\documentclass[a4paper,twocolumn,11pt,unpublished]{quantumarticle}
\pdfoutput=1
\usepackage[utf8]{inputenc}
\usepackage[english]{babel}
\usepackage[T1]{fontenc}
\usepackage{amsmath}
\usepackage{hyperref}
\usepackage{fixltx2e}
\usepackage[numbers,sort&compress]{natbib}

\usepackage{tabularray}
\usepackage{booktabs, makecell, multirow,}
\usepackage[export]{adjustbox}

\usepackage[bb=dsserif]{mathalpha}
\usepackage{bm}

\usepackage{amssymb}
\usepackage[linesnumbered,lined,ruled]{algorithm2e}

\DeclareMathOperator*{\argkmax}{arg~k-max} 

\begin{document}

\title{Discovering Quantum Circuit Components with Program Synthesis}

\author{Leopoldo Sarra}
\affiliation{Max Planck Institute for the Science of Light, 91058 Erlangen, Germany}
\affiliation{Department of Physics, Friedrich-Alexander Universität Erlangen-Nürnberg, 91058 Erlangen, Germany}
\orcid{0000-0001-7504-8656}
\author{Kevin Ellis}
\affiliation{Cornell University, United States}
\orcid{0000-0001-6586-0632}
\author{Florian Marquardt}
\affiliation{Max Planck Institute for the Science of Light, 91058 Erlangen, Germany}
\affiliation{Department of Physics, Friedrich-Alexander Universität Erlangen-Nürnberg, 91058 Erlangen, Germany}
\orcid{0000-0003-4566-1753}

\maketitle

\begin{abstract}
    Despite rapid progress in the field, it is still challenging to discover new ways to take advantage of quantum computation: all quantum algorithms need to be designed by hand, and quantum mechanics is notoriously counterintuitive.
    In this paper, we study how artificial intelligence, in the form of program synthesis, may help to overcome some of these difficulties, by showing how a computer can incrementally learn concepts relevant for quantum circuit synthesis with experience, and reuse them in unseen tasks.
    In particular, we focus on the decomposition of unitary matrices into quantum circuits, and we show how, starting from a set of elementary gates, we can automatically discover a library of new useful composite gates and use them to decompose more and more complicated unitaries.
\end{abstract}

\begin{figure}
    \centering
    \includegraphics[width=\the\columnwidth]{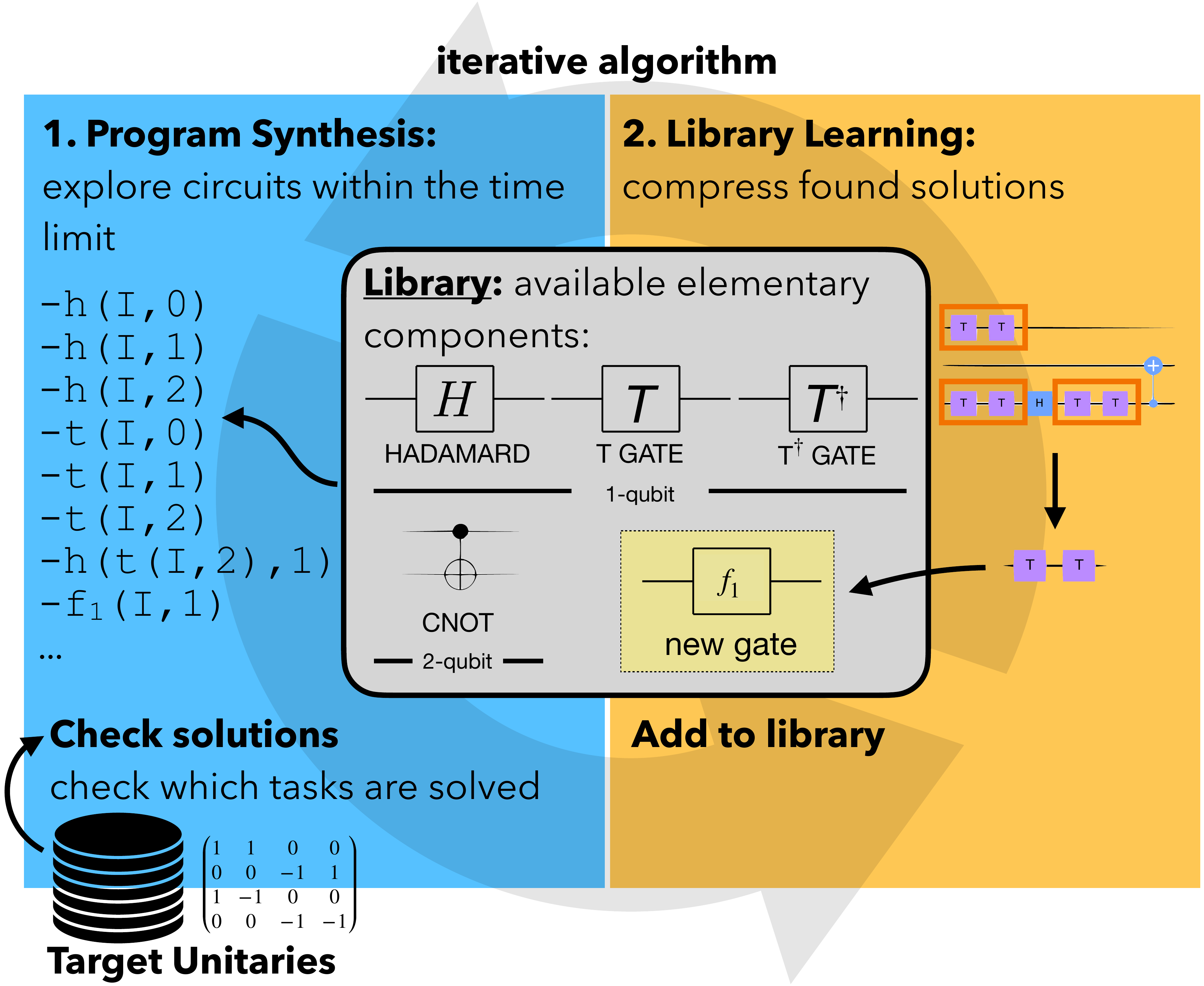}
    \caption{Synthesis of quantum unitaries.
        Given a dataset of target unitary matrices, we enumerate quantum circuits for a given timeout interval, using as components the elementary gates in the library.
        After some matrix decompositions have been found, the solutions are analyzed and the most useful components are added to the library as new available gates.
        The procedure is repeated for a given number of iterations.}
    \label{fig:algorithm}
\end{figure}
It has been theorized for decades that quantum computers can perform tasks more efficiently than classical ones~\cite{feynman_simulating_1982}.
Consider for example Shor's algorithm for factorization~\cite{shor_algorithms_1994} or Grover's algorithm for search ~\cite{grover_fast_1996}, the variational quantum eigensolver~\cite{peruzzo_variational_2014} or other quantum machine learning algorithms~\cite{biamonte_quantum_2017}, which can all have a large impact on important problems.
Even though those algorithms are already very promising, and they justify the current effort in the development of large-scale quantum computers~\cite{preskill_quantum_2021}, it is hard to automatically exploit the advantages offered by quantum computing.
Indeed, each of those algorithms has been invented specifically for the task they solve, and often their principles do not easily generalize to other tasks.
Up to now, we can barely count more than two hundred algorithms in total~\cite{montanaro_quantum_2016}.
While there is strong evidence of a quantum advantage~\cite{preskill_quantum_2012}, i.e. quantum computers can be more powerful than classical ones, and this advantage on near-term devices has been even shown experimentally on specific purpose-designed tasks~\cite{aaronson_computational_2011, arute_quantum_2019, zhong_quantum_2020}, we do not have a way to automatically make use of quantum principles like superposition or entanglement to speed up classical algorithms: each algorithm must be designed from scratch, and it is not clear a priori whether a corresponding faster quantum algorithm exists.
Compared to the classical regime, quantum mechanics is generally counterintuitive to human common sense, and thus scientists require a large effort and imagination to conceive quantum algorithms.
Hence, it would be useful if there was a technique that helped to understand the laws of the quantum world and aided in finding an efficient way to solve a given task on a quantum computer.
It would be an invaluable tool to assist researchers in the development of new quantum algorithms, perhaps contributing to understanding general principles that can grant quantum speed up.
This long-standing goal, if attainable at all, is clearly out of reach at the moment, but it is interesting to explore it and develop ingredients that could be useful for that purpose.

This paper takes a step in that direction by considering a simpler yet core subproblem:
instead of working on entire quantum algorithms, we focus on the purely quantum part, excluding measurements and classical processing.
We develop a machine learning technique to automatically produce a quantum circuit that performs a requested unitary transformation of a quantum state. The main innovation of our approach to this general problem is the gradual discovery of new composite gates, which can be subsequently used to decompose more and more complicated unitary matrices.
In this way, we can build a self-learning compiler, which can express a given unitary matrix into a given set of gates and respect a given qubit connectivity, without ever providing explicit transformation rules.
Instead, we just propose a training set of unitary matrices to decompose, with varying degrees of difficulty.

At a high level, our method works by iteratively
(1) searching for correct decompositions, assembling together gates from our current set of components, and
(2) extracting \emph{new} components by analyzing the solutions found in the previous step.
In doing so, our system gradually learns increasingly useful quantum operations, which it uses to decompose more and more complex matrices (Fig.~\ref{fig:algorithm}).
At first, our method automatically rediscovers suitable representations of common gates like SWAP and CZ, even when these gates are not initially provided to the system.
Importantly, it also proposes new unexpected combinations of gates that prove to be valuable ingredients in the construction of more sophisticated unitaries.
Circuits that would be too difficult to synthesize by randomly assembling the initial elementary components can be tractably found using new gates that the system itself proposed to use when building easier circuits.
In that way, the system bootstraps a set of new gates from solving easier problems, which unlocks solving harder problems, which leads to new gates, and so on.

The technique here presented introduces the ideas of concept extraction and program synthesis to the field of quantum computing and shows proof-of-concept applications to the domain of unitary matrix decomposition.
We show the importance of bootstrapping more advanced concepts from simpler ones, and how these concepts can then be used to solve increasingly complex problems.
The choice of expressing concepts as small "programs" also allows for better interpretability, in contrast to other possible black-box approaches like neural networks~\cite{goodfellow_deep_2016}.
Extensions of these ideas will allow significant improvements to existing machine learning methods in quantum computing.

\section{Related Works}
The synthesis of quantum unitary matrices, to which we will also just refer as "unitaries" in the following, is the process of building a quantum circuit by placing gates one after the other, acting on selected qubits, to reproduce the effect of the given unitary.
The problem of building a circuit that reproduces the action of a given unitary, or of another quantum circuit, using only a given set of gates, or respecting some given constraints, is well-known in the literature, and is called compilation~\cite{benenti_principles_2004}.
On the other hand, if the unitary is given by a circuit expressed in a set of gates, and we want to find an equivalent circuit which uses another set of gates, the process is called transpilation.
Many algorithms already exist to compile given unitaries into circuits~\cite{ostaszewski_structure_2021} that only use a given set of gates, e.g. the ones that can be physically implemented, some even using machine learning and reinforcement learning techniques~\cite{moro_quantum_2021}.
Many other works explore different directions for optimizing the total number of gates in a circuit~\cite{fosel_quantum_2021} or reducing the number of a given kind of gate, which may be more expensive or noisy in the considered implementation~\cite{wang_automated_2022}.
For example, the Solovay-Kitaev algorithm~\cite{kitaev_quantum_1997, dawson_solovay-kitaev_2006} can approximate with arbitrary error any given unitary matrix with a given finite set of gates, as long as this set can approximate any one-qubit gate.
In our case, we build a self-learning compiler, which changes its behavior according to the experience, until converging to the optimal solution for the given architecture.
This kind of compiler does not need explicit rules about how to decompose the given unitary matrix, but just a series of matrices to decompose with increasing difficulty.
We emphasize that the main interest in our case is not in reproducing the performance of a state-of-the-art compiler, or transpiler, but to investigate how to imitate the ability of scientists to learn new concepts and reason with them, for example by building more and more complicated circuits from elementary components whose behavior is known.
In our examples, we mainly optimize for conceptual efficiency, i.e. number of used high-level operations to express a quantum operation, rather than for the effective experimental implementation cost (e.g. total gate number minimization). Different constraints can potentially be chosen nonetheless.

After expressing quantum circuits more conveniently, we can take advantage of machine learning techniques to assemble them until the requested unitary is produced.
We build on machine learning techniques for program synthesis~\cite{gulwani_program_2017}.
Program synthesis methods automatically construct source code, and our work exploits the fact that a quantum circuit can be easily expressed as a simple program.
Recent program synthesis techniques use neural networks to learn how to generate source code (\cite{li_competition-level_2022}, \emph{inter alia}).
Our work uses a slightly different family of learning methods which casts program synthesis as Bayesian inference~\cite{saad_bayesian_2019}.
This probabilistic Bayesian framing allows searching for the most likely programs that solve a given unitary, and also learning to generate good programs via hierarchical Bayesian methods.
These Bayesian program learning methods were developed in a series of works~\cite{dechter_bootstrap_2013,liang_learning_2010,ellis_library_2018}.
We directly build upon DreamCoder~\cite{ellis_dreamcoder_2021}, a recent work in this family.

Coming back to the much higher conceptual level, regarding the longer-term idea of an "artificial scientist", the first proof-of-principle of an agent capable of conducting research on its own has been shown in~\cite{king_automation_2009}, to automate functional genomics experiments.
The idea of automatic concept discovery from experience is a helpful ingredient for this long-term goal, as it is a first step towards reasoning.
Indeed, there has already been a large interest also in other fields of physics ranging from the design of new quantum optics setups~\cite{arlt_digital_2022}, the use of symbolic regression and graph neural networks to find new laws of astrophysics~\cite{cranmer_discovering_2020}, to the general development  of algorithms capable to formulate scientific laws~\cite{wu_toward_2019}.
In~\cite{trenkwalder_automated_2022}, the idea of extraction of building blocks is used in a reinforcement learning setting to produce quantum entangled states.
Also, projective simulation~\cite{briegel_projective_2012} allows employing reinforcement learning agents to explore novel algorithms for quantum communication~\cite{wallnofer_machine_2020}.
While these techniques share the idea of concept extraction to solve a specific quantum task, in the usual reinforcement learning setting the discovery of new components is an incidental effect, obtained while achieving the task.
In our case, the goal of circuit decomposition itself is the discovery of new useful gates, by minimizing the overall complexity of the solutions, quantified in terms of description length~\cite{rissanen_modeling_1978, bishop_pattern_2006}.

\section{Methods}
\begin{figure}
    \centering
    \includegraphics[width=\the\columnwidth]{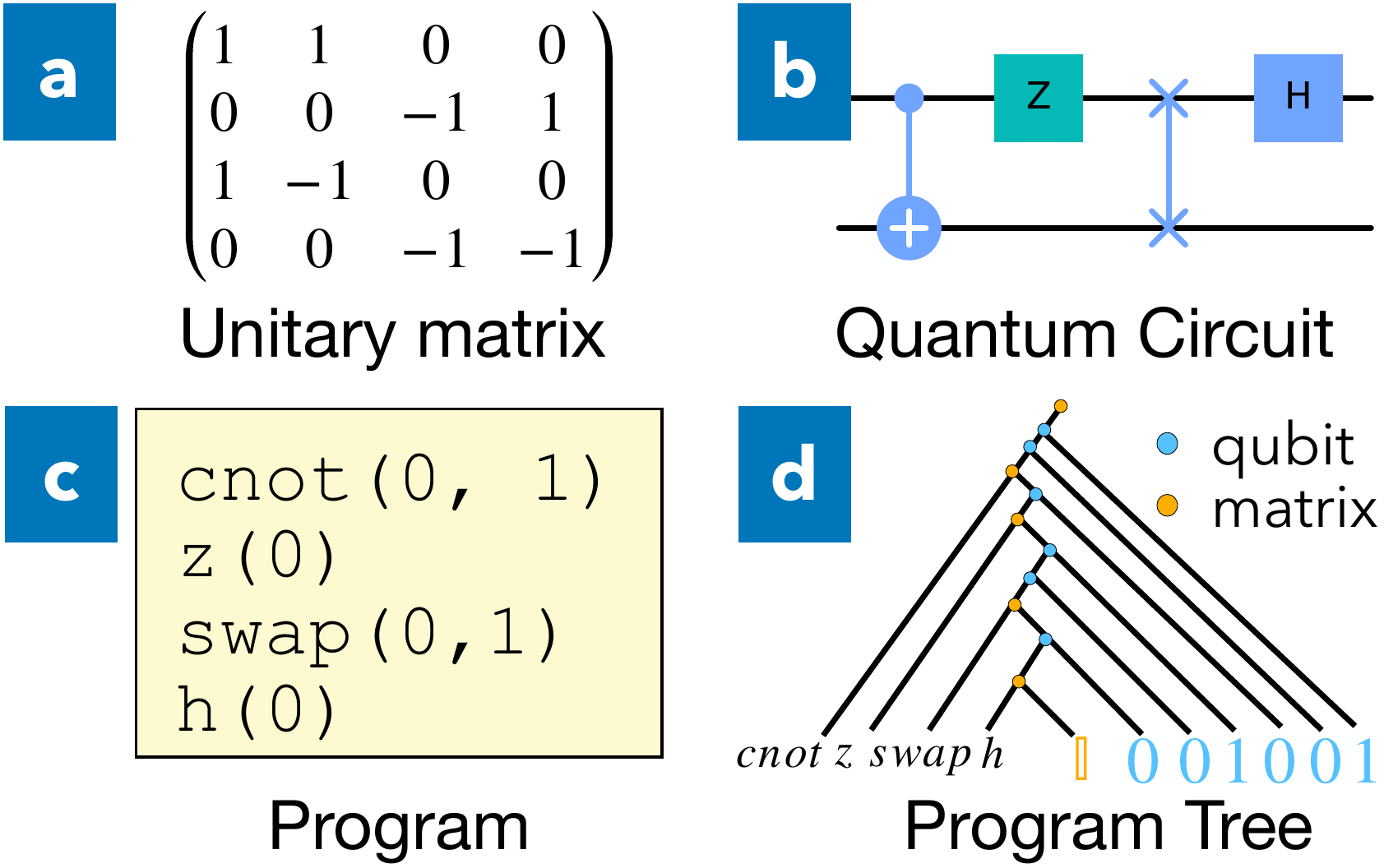}
    \caption{Unitary matrix representations.
        A quantum unitary matrix (a) can be represented as a quantum circuit (b), a sequence of instructions about the gates and the qubits on which they act (c), or a tree representation of a program (d).
        When enumerating new programs, the input identity matrix ($\mathbb I$) is replaced with another gate applied to other qubits, creating a new branch. }
    \label{fig:unitaries}
\end{figure}

In this section, we explain how our algorithm for unitary synthesis and gate extraction works.
As shown in Fig. \ref{fig:unitaries}, quantum circuits can be seen as programs that subsequently apply operations to a given state: starting from the identity matrix, representing a circuit without any gates, each operation corresponds to a multiplication by the unitary matrix associated to the applied gate.
The final unitary matrix is built up by sequentially multiplying all the unitary matrices.
By working on programs that build up the sequence of operations that make a circuit, we can explore  the space of possible unitary matrix decompositions with program synthesis.

To begin, we define a probability distribution over quantum circuits $c$.
Our learning algorithm works by adjusting the distribution over $c$ to make useful circuits more likely.
The distribution depends on the set of allowed gates, $G$, together with the probability of each gate $g\in G$, which we write $\theta_g$.
We decompose it as a product of the probability of choosing the specific gates and applying those gates to the selected qubits.
We assume gates are generated independently at random, and that they attach to wires (i.e. qubits) drawn uniformly and independently at random:
\begin{align}
    \label{eq:circuit-probability}
    P(c|G, \theta) & =\prod_{g\in c} \mathbbb{1}\left[ g\in G \right] \theta_g \chi(c,g),
\end{align}
where $\mathbbb{1}$ is the indicator function, yielding one iff the condition is fulfilled,
and $\chi(c,g)$ the probability of connecting the gate to the specific qubits.
For example, we use uniform probability of associating a gate to any of the possible qubits: $\chi(c,g) =N_c^{-n_g}$, with
$n_g$ is the number of inputs to gate $g$,
$N_c$ is the number of qubits in the circuit $c$.
If the resulting circuit is not valid, for example if the inputs of a CNOT gate are repeated, it is automatically discarded, which leads to a small modification of the probability $\chi$ whose discussion we omit since it is not crucial for understanding the workings of the algorithm.

Notice that by optimizing the choice of the set of gates $G$, it is possible to make more complex circuits more likely.

Ultimately our aim is not to probabilistically score specific circuits but to learn a collection of gates that are valuable for solving a broad range of unitary synthesis problems.
To that end, we assume that we have a training set of unitary matrices to decompose, collectively written $U$.
Our algorithm tries to maximize the posterior probability of the gate set, given that it must solve every unitary in $U$: it finds the optimal gate set and optimal gate probabilities as
\begin{align}
    \label{eq:optimization-goal}
    (G^* , \theta^*) = \arg \max_{G,\theta} P(G, \theta |U),
\end{align}
where we employ Bayesian reasoning to write
\begin{align}
     & P(G, \theta|U)\propto                                                                \\
     & \;\;P(G)P(\theta|G)\prod_{u\in U}\sum_{c}\mathbbb{1}[\mathcal{U}(c)=u]P(c|G,\theta),
    \label{eq:posterior}
\end{align}
where $P(G)$ is the prior over gate sets, $P(\theta|G)$ is the prior of gate weights of a given gate set, and $\mathcal{U}(c)$ is the operator that gives the unitary matrix associated to a given circuit.
The above equation is computationally intractable because it includes summing over the infinite space of all circuits (inner sum over $c$).
We introduce a tractable lower bound on Eq.~\ref{eq:posterior} by only summing over a small set of possible circuits for each unitary.
Writing $\mathcal{B}_u$ for the small set of circuits we consider for unitary $u$, our objective function becomes lower-bounded by
\begin{align}
     & P(G)
    P(\theta|G)\prod_{u\in U}\sum_{c\in \mathcal{B}_u}\mathbbb{1}[\mathcal{U}(c)=u]P(c|G,\theta).
    \label{eq:lowerbound}
\end{align}
Eq.~\ref{eq:lowerbound} serves as our core objective function for learning a library of gates. A more detailed derivation of this objective is given in the Appendix~\ref{app:probabilistic-framing}, and in the original work \cite{ellis_dreamcoder_2021}.
Maximizing it with respect to $(G,\theta)$ corresponds to updating our gate set to increase the probability of a circuit solving each unitary.
Maximizing it with respect to $\mathcal{B}_u$ corresponds to program synthesis: finding a handful of likely  circuits that evaluate to a given unitary.

More precisely, our system takes as inputs the example unitary matrices to learn to decompose, $U$, together with a set of initial elementary gates, $G_0$.
The unitaries provided as examples in $U$ determine which assemblies of gates are the most useful, thus the optimal set of learned gates $G^*$.
The algorithm iterates many times through two phases:
program synthesis, where circuits that decompose target matrices are proposed, and library learning, where concepts are extracted from the found circuits and the most useful sequences of gates are added to the set of elementary gates $G$ as a composite gate. It can then use the new gate as a single block in the subsequent iterations of program synthesis.
Mathematically, program synthesis corresponds to maximizing Eq.~\ref{eq:lowerbound} w.r.t. $\mathcal{B}_u$, while library learning corresponds to maximizing w.r.t. $(\theta,G)$.
A sketch of the algorithm is shown in Fig.~\ref{fig:algorithm}.

\subsection{Program Synthesis}
During this phase of the algorithm, we seek the top $k$ most likely programs solving each unitary:
\begin{align}
    \mathcal{B}_u\gets \argkmax_{c} P(c|G,\theta) \mathbbb{1}[\mathcal{U}(c)=u],
\end{align}
where $\argkmax_c$ is the function that returns the arguments with the largest $k$ values.
To find those top $k$ circuits, we enumerate programs in order of decreasing probability under $P(\cdot |G,\theta)$ until $k$ solutions have been found or we reach a timeout.
We construct the syntax trees of candidate programs bottom-up, with higher probability expressions being generated first, using recent algorithms for probabilistic program enumeration~\cite{barke_just--time_2020,fijalkow_scaling_2022}.
As an optimization, we discard any programs containing subexpressions that are semantically equivalent to higher-probability subexpressions, meaning they evaluate to the same unitary (in the literature called pruning by observational equivalence~\cite{udupa_transit_2013}).
Within our implementation, we search for a maximum budget of $200$ seconds and collect the top $k=2$ programs for each unitary.

In practice, to accelerate the convergence in the algorithm, we do not update the programs for every unitary at each iteration.
Instead, we sample a small batch of unitaries and only synthesize programs for those tasks.
This is analogous to the use of mini batching for training neural networks using gradient descent~\cite{goodfellow_deep_2016}.
Essentially, it allows taking fast, small learning steps (updating the set of available gates $G$) without examining and analyzing the entire training set.

Although enumeration may seem like a very basic program synthesis strategy, our goal is to learn a sophisticated set of gates $G$  such that even a simple enumerative search can quickly uncover interesting unitaries.
Thus, the ability of the program synthesis to succeed hinges critically on learning a good gate library $G$, which we describe next.

\subsection{Library building}
During the library building step, we augment the library of gates by adding new compositions of gates that the system itself proposes.
We do this by analyzing the circuits found during the program synthesis phase and extracting commonly occurring patterns of gates.
Adding these new patterns of gates to $G$ increases the probability of generating circuits that use them.

On the other hand, the goal of library building is not to simply memorize every successful circuit, even though memorizing would most increase the probability of the programs found so far.
Instead, we want to find new gates that generalize the patterns found in the synthesized programs.
Striking the right balance is accomplished by prioritizing gate sets that are compressive, i.e. have small description length~\cite{rissanen_modeling_1978, bishop_pattern_2006}.
Remembering that our goal is to maximize Eq.~\ref{eq:posterior}, we see that we need to not just make the circuits likely under $G$, but also have a $G$ with a high prior probability $P(G)$.
Our system uses a prior that assigns less probability to larger sets of gates and to gates with many subcomponents, which exerts pressure for proposing new gates that are small, yet broadly useful across many tasks.

Algorithmically, our system proposes new gates by extracting fragments of program syntax trees discovered during the previous program synthesis phase.
Given a set of candidate new gates, $G'$,
it then constructs a set of candidate new libraries that extend the old library by exactly one gate: $\left\{ G\cup\left\{ g' \right\}\;:\;g'\in G' \right\}$.
For each such $G\cup\left\{ g' \right\}$, the system estimates a new $\theta$ using Expectation-Maximization~\cite{bishop_pattern_2006}.
The system finally computes the objective function in
Eq.~\ref{eq:lowerbound}, and takes the gate which most increases it.
This entire process repeats until Eq.~\ref{eq:lowerbound} fails to improve, and then another round of program synthesis begins.
See~\cite{ellis_dreamcoder_2021} for details.

Despite the apparent simplicity of the synthesis step, the overall algorithm is substantially more efficient than simple brute-force enumeration, as each component is used according to its assigned probability, and branches of the tree are pruned as they are discovered to be equivalent to already known branches.
Also, the addition of the extracted gates to the set of elementary gates increases the breadth of the search tree (more components to choose from at each step) but reduces the required depth of the search (number of components to put together one after the other).
Without these tricks, there would be a combinatorial explosion with the depth of the tree (e.g. $\mathcal{O}((gn)^d)$ with $g$ elementary gates, $n$ qubits and depth of the tree $d$, considering only 1-qubit gates in this rough estimate).
It would be just unfeasible to decompose very long circuits, and this is why reducing the depth to explore is so helpful.
Using the probabilistic guidance of the learned $(G,\theta)$, we can discover a circuit $c$ in at most $\mathcal{O}(1/P(c|G,\theta))$, which may be much better than $\mathcal{O}((gn)^d)$ if the target circuit $c$ employs similar computational motifs to the training data.
In some sense, this algorithm allows us to learn a domain-specific language for quantum circuits, by discovering a good prior to guide the circuit synthesis.

\section{Results and discussion}
\begin{figure}
    \centering
    \includegraphics[width=\the\columnwidth]{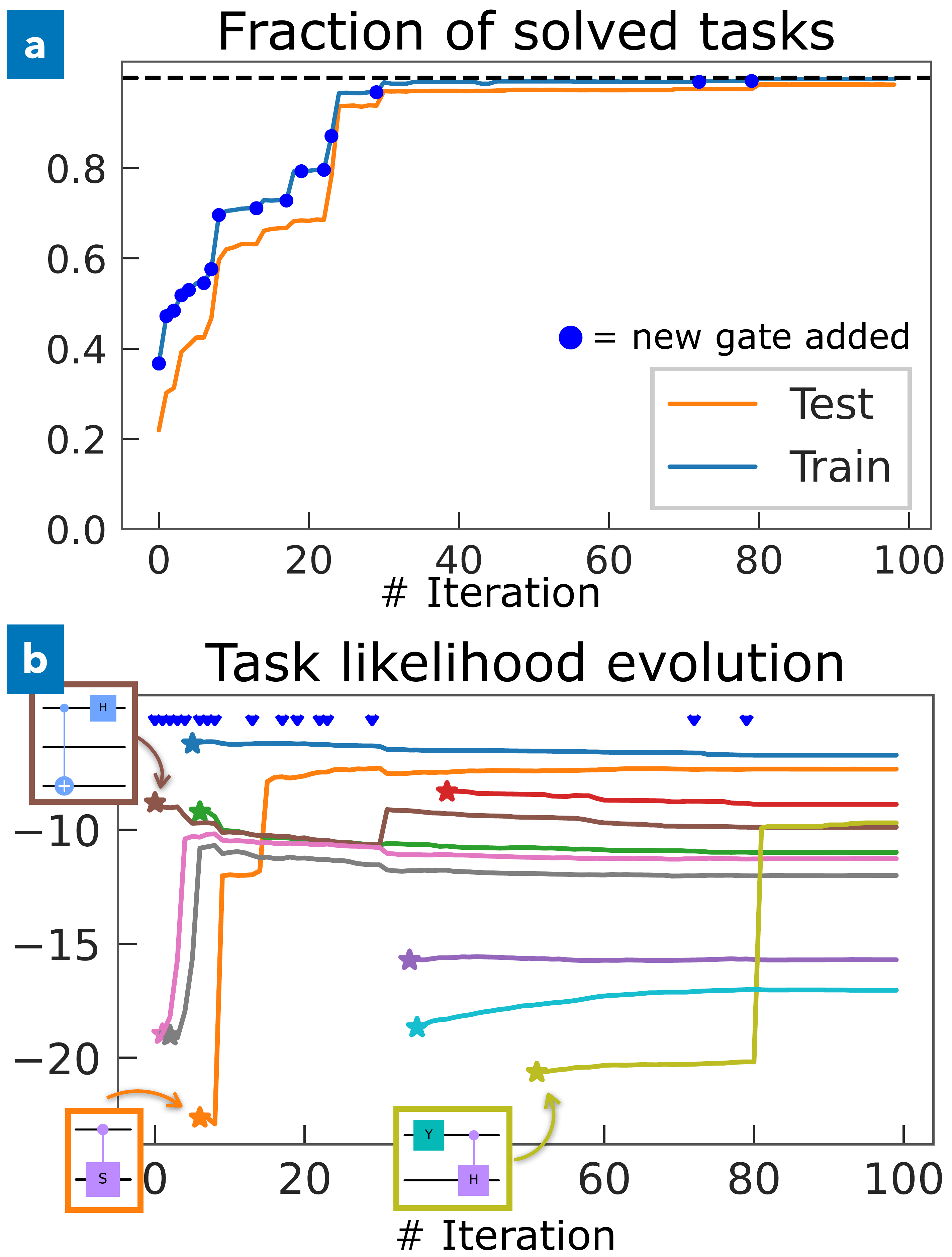}
    \caption{Unitary synthesis with full connectivity among qubits.
        (a) Fraction of solved tasks at each iteration.
        (b) Likelihood to decompose some target unitaries during the algorithm iterations ($P(u|G,\theta)$).
        The star symbol shows the first iteration in which the target matrix has been successfully decomposed.
        As new gates are discovered, some matrices become easier to decompose.
        For three tasks, we show the circuit that generates the target matrix in the insets.
        In both figures, the blue symbols mark the iterations at which a new gate has been extracted.}
    \label{fig:fully-connected}
\end{figure}

In this section, we show the application of our unitary matrix decomposition algorithm using an elementary set of gates, which can theoretically approximate any circuit.
We show results when enforcing either full connectivity between qubits, or only allowing gates between nearest-neighboring qubits (e.g. between qubit $0$ and $1$ but not $0$ and $2$).
To make the search faster and focus on the proof of concept, we limit ourselves to discrete gates, i.e. gates that do not depend on a tunable real parameter: this would require an optimization over the parameter of the gates, in addition to the search among the possible programs.
For simplicity, we also fix the number of qubits to be the same in the entire set.
In particular, to avoid the combinatorial explosion due to input qubit combinations (a $n$-qubit gate should be tested on all permutations of qubit inputs), we limit our examples to circuits with only $3$ qubits, but generalizations to larger circuits are of course possible.

We choose $G_0=$\{H, T, T$^\dagger$, CNOT\} as the elementary gate set.
This essentially corresponds to the Clifford gate set, plus T gates to make it a universal approximator~\cite{ kliuchnikov_fast_2012, giles_exact_2013}.
We also include the $T^\dagger$ gate, which itself corresponds to seven $T$ gates, to spifeed up the search.
The choice of the target unitary set $U$ is important, as the extracted gates will be selected to maximize the decomposition efficiency over those tasks.
In general, some unitaries are much harder to approximate because they require many elementary gates.
Sampling from the space of unitary matrices would generally produce matrices that are too hard to decompose when starting from our elementary set of gates and trying combinations of them.
In our experiments, to be sure that it can be decomposed in a finite amount of search time, we build $U$ by defining another set of gates, $G_\text{tasks}$, which uses more high-level operations.
We sample circuits from  $G_\text{tasks}$ by randomly putting gates on the circuit.
The unitaries associated with the sampled circuits will be the target for our algorithm.
To keep the decomposition difficulty under control, also the gates in  $G_\text{tasks}$ have no continuous parameters, in particular $G_\text{tasks}=\{$H, T, T$^\dagger$, S, X, Y, Z, SX, SX$^\dagger$, CNOT, CY, CZ, CS, CH, SWAP, iSWAP$\}$.
We first generate a set of matrices by enumerating all the possible circuits given by this set within $50$ seconds.
From this set, we select $1000$ matrices to build $U$.
To make sure that the train set always contains a significant fraction of both easy and difficult decomposition tasks, we choose them with uniform probability in the number of gates of the initial circuits, taking into account only circuits generated within the timeout.
All other generated unitaries are included in our test set $T$, and they will be used to assess the performance of the algorithm.
The target dataset thus contains tasks with different levels of difficulty, allowing the algorithm to gradually learn to solve more and more complicated tasks.
It is important to include unitaries with different decomposition lengths.
Indeed, only after some tasks are solved it is possible to extract gates that can be used to solve other tasks since we need at least some elements in $B_k$ in Eq.~\ref{eq:lowerbound}.
If all tasks are too complicated, the learning procedure will not start and each iteration will not provide any benefit, since it will always propose the same programs.
In that case, the enumeration timeout should be increased until some solutions are found.

\begin{figure}
    \centering
    \includegraphics[width=\the\columnwidth]{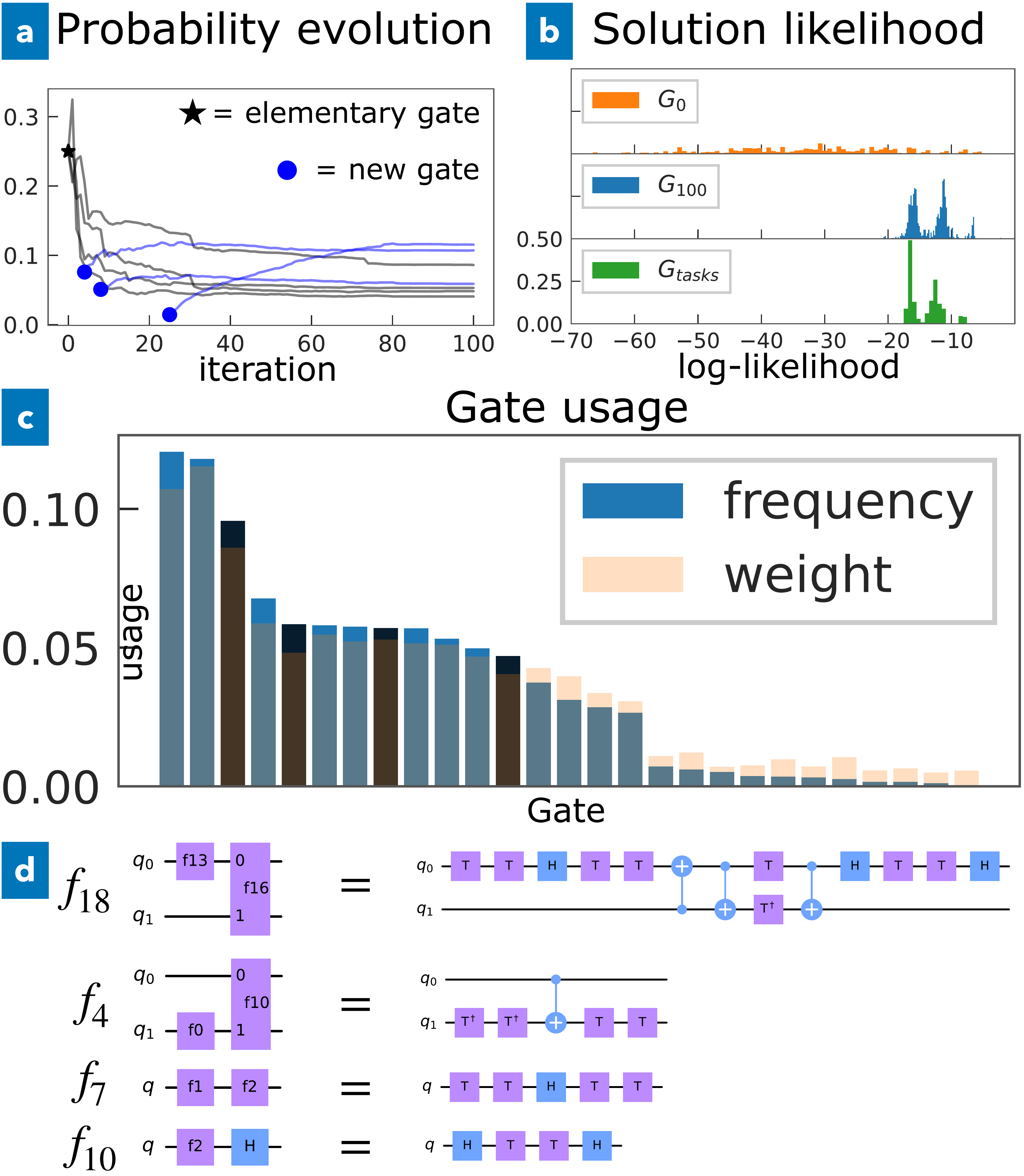}
    \caption{Analysis of extracted gates.
        (a) Evolution of the probability of using a certain gate ($\theta$).
        Some extracted gates (in blue) become more important during the iterations, while some elementary gates (in black) become rarer.
        (b) Likelihood of decomposing the target unitaries with different elementary sets: the initial set of gates, the final one (which also includes the extracted gates), and the set of gates with which the target dataset has been generated.
        (c) Final probability of using a certain gate in $G_{100}$ ("weight") and in the found solutions ("frequency").
        In blue are shown the extracted gates, in black the initial ones (in decreasing order: CNOT, T, Hadamard, and T$^\dagger$).
        (d) The first few most useful extracted gates.
        \label{fig:primitive-analysis}}
\end{figure}
We perform $100$ algorithm iterations, each time considering batches of $25$ tasks and enumerating for $150$ seconds (in parallel with $32$ CPUs).
We see that after some iterations we can solve most of the about $50 000$ of tasks in the test set.
Every time we learn a new gate, the algorithm starts using it to explore new circuits thus solving more tasks.
To evaluate the performance of the algorithm, we can consider the test set $T$ (which the algorithm has never seen during the previous iterations) and check how many unitaries can be decomposed into a circuit.
Results are shown in Fig.~\ref{fig:fully-connected}. After about $50$ iterations, the algorithm can decompose almost all the proposed matrices.
The algorithm rediscovers suitable decompositions of gates into the available elementary gates.
For example, it finds useful elementary decompositions of high-level gates like the SWAP gate and the Pauli gates, adding them as building blocks to its library. Importantly, it goes beyond those simpler examples and discovers more complicated composite gates, which also seem to be useful to reuse.
By discovering useful building blocks, matrices that are initially impossible to decompose in the given time budget because of the high number of required components are easily decomposed into short sequences of the newly extracted components.
Indeed, as soon as a new gate is extracted, many new unitary matrices can immediately be decomposed.
The "train" and "test" curves are obtained by checking all the found programs at a given iteration against each unitary matrix in the target set $U$ and test set $T$.
We recall that here we consider the whole target set for evaluation purposes, but the algorithm only has access to a small random batch of matrices at each iteration.
As Fig.~\ref{fig:fully-connected}b shows, the difficulty to decompose a given matrix changes during the algorithm iterations: when new gates are added, some unitaries suddenly become easier to be decomposed, while other matrices, which mainly use the initial elementary components, become less likely to occur because those components become less frequent in the enumeration.
Overall, all matrices become easier to decompose.

It is also interesting to inspect the library of extracted gates and try to interpret their behavior.
As seen in Fig.~\ref{fig:primitive-analysis}, after discovering composite gates, initial gates like T become less useful and are used less often, while some new gates like "f$18$", which corresponds to a controlled Hadamard gate suitably decomposed into elementary gates, and "f$4$", which corresponds to a controlled Z gate, are used more often than $CNOT$.
By looking at Fig.~\ref{fig:primitive-analysis}b, we see that, initially, it is generally very hard to find decompositions for the target matrices (in orange).
After running the algorithm, we have a new set of gates (in blue) that allows decomposing most of the matrices.
We go from problems with probability~$\simeq e^{-60}$ to be solved to~$\simeq e^{-20}$, about $17$ orders of magnitude larger probability, which make it feasible to find the decomposition in a finite amount of time.
It is interesting to observe how a different choice of elementary components can make the decomposition easier.
In particular, the extracted set of gates at the final iteration, $G_{100}$, makes the decomposition of the target matrices even easier than when using exactly the same set we used to generate the target dataset itself (in green).
In other words, our algorithm discovers a set of quantum gates to describe the target dataset that is even better than the set that we used to generate it, $G_\text{tasks}$.
For example, it turns out that it is much more useful to have two-qubit gates like $CH$ and $CZ$ than CNOT.

\begin{figure}
    \centering
    \includegraphics[width=\the\columnwidth]{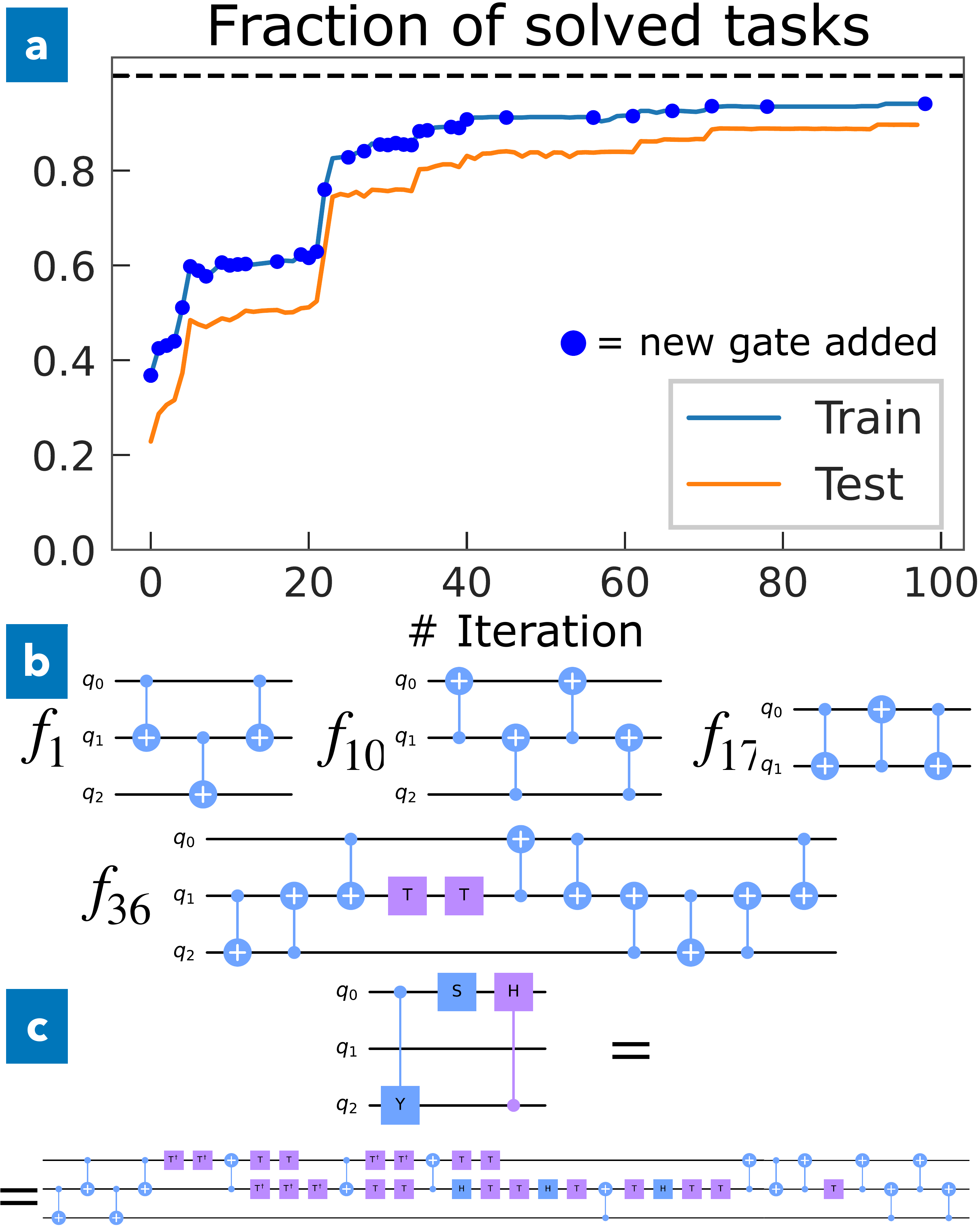}
    \caption{Unitary synthesis with only nearest-neighbor connectivity among qubits.
        (a) Percentage of solved tasks at each iteration.
        (b) Some extracted gates.
        (c) Example decomposition of a high-level circuit into the requested gate set.}
    \label{fig:limited_train}
\end{figure}

We also performed another experiment with the same parameters, but this time we constrained the 2-qubit gates to only act on neighboring qubits.
This configuration resembles a linear array of qubits, where interactions are constrained to nearest neighbors.
Also in this case, the algorithm learns to decompose more and more matrices with experience.
We notice that this time a larger fraction cannot be decomposed yet even after $100$ iterations.
This is due to the larger difficulty of this problem, and with more iterations and larger enumeration timeout results would keep improving.
Again, the algorithm learns more complicated gates and finds similar results as in the previous example.
In addition, it also learns gates that allow it to efficiently handle the enforced connectivity constraint, like the SWAP gate between the first and the third qubit (by swapping with the middle one) and similar two and three-qubit gates.
The automatic extraction of composite gates allows for expressing concisely very long sequences of gates.
Results are shown in Fig.~\ref{fig:limited_train}.

The final outcome of the algorithm depends of course on the chosen definition of being a better set of gates: in this case, we wanted to minimize the number of components to put in a circuit so that all target matrices could be decomposed with some high-level gates.
However, different constraints can be considered, for example, to include the overall length of the circuit in terms of elementary components or a different cost for the use of each component.

\section{Outlook}

In this paper, we have shown how concept extraction and program synthesis techniques can potentially help quantum computing, by providing tools to work and reason with the quantum world.
In particular, we have shown a procedure to discover useful quantum gates (in terms of reusability) by just giving a set of unitary matrices to decompose.
This can be seen as a first step toward the longer-term goal of enabling the discovery of new quantum algorithms.

The extension to larger qubit numbers will require careful optimization of the performance of the different parts of the algorithm (search and library building), but we anticipate there is a lot of room for improvement here.
Indeed, our experiments take about $20$ minutes per iteration to run, and systems with more qubits would require much more time.
It is also possible to test the generalization capabilities by running first on smaller systems and then trying to solve more complicated tasks on larger systems.
To improve performance, it would be possible to restrict the decomposition to the Clifford gate set, so that calculation would be faster, e.g. by exploiting highly optimized Clifford simulators that can deal with large qubit numbers~\cite{gidney_stim_2021}.
To tackle the combinatorial explosion due to the larger number of possible qubits a gate can be applied to, more advanced approximations for the circuit distribution $P(c)$ in Eq.~(\ref{eq:circuit-probability}) may be employed.
For example, instead of factorizing the circuit distribution as the product of the probability of its gates, we could condition the probability of a gate to the previous ones, improving the precision of the enumerator.

One of the most important future extensions would concern the choice of the set of target unitaries.
While in our case these were generated as random circuits from a high-level gate set, the application to more structured training sets would greatly increase the power of the approach.
We are thinking, in particular, of circuits generated from a library of quantum algorithms.

Different connectivity constraints may also be enforced, or one could extend the programs that generate the circuits to also include programming constructs like conditions and loops.
In the long run, the presented library bootstrapping procedure can be part of future algorithms to automatically extract components and reuse them in a curriculum-learning approach~\cite{wang_survey_2021}.
Also, it would be possible to adopt this concept extraction algorithm as an additional step of a reinforcement learning agent \cite{sutton_reinforcement_2018} that tries to decompose a unitary matrix.
In that case, the goal would be to train an agent (i.e. a probability distribution of putting a certain gate given the current circuit) to synthesize the unitary, where the state would be the current circuit, and the possible actions would be the allowed elementary gates.
It would be possible to extend the action space by adding the extracted gates, thus facilitating the exploration of the circuit space, as in hierarchical reinforcement learning \cite{pateria_hierarchical_2022}.

Finally, on a more general level, the ability to extract concepts and to use them in further exploration is also interesting for the development of future "artificial scientist" algorithms, here  applied to the quantum domain and specifically quantum computation, aimed at reasoning and developing scientific models similarly to humans: the possibility to define concepts and reason about them is reasonably a necessary skill for this purpose.
Similar techniques can be a useful addition to existing machine learning algorithms for quantum circuit design, and, in the long run, they may help to develop new quantum algorithms.

The code of the algorithm and the instructions to reproduce the presented examples are open-sourced on GitHub
\footnote{\url{https://github.com/ellisk42/ec/tree/quantum_algorithms-simplified}}.

\section*{Acknowledgments}
This work was supported by the Munich Quantum Valley, which is supported by the Bavarian state government with funds from the Hightech Agenda Bayern Plus, and by the Max Planck Society.


\bibliographystyle{quantum}
\bibliography{bibliography/UnitarySynthesis}

\onecolumn
\clearpage
\appendix
\section{The program synthesis algorithm}
To apply the program synthesis framework presented in \cite{ellis_dreamcoder_2021}, we need to express quantum circuits as programs.
The specific formalism that we adopt is that of functional programming, typed-$\lambda$-calculus~\cite{pierce_types_2002} in particular.
Each quantum gate becomes a function that takes as input a quantum circuit and the sequence of qubits on which it should act, and returns a quantum circuit with the requested gate applied on the right.
In this way, a program is simply the sequential application of many gates, starting from the empty circuit $\mathbb{I}$.
\begin{figure}
    \centering
    \includegraphics[height=3cm]{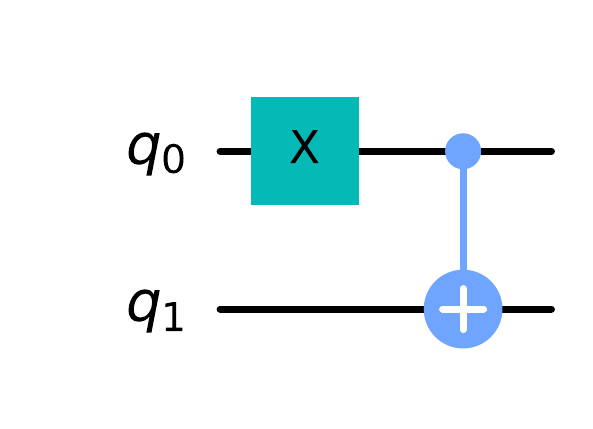}
    \caption{Example circuit}
    \label{fig:example-cnot}
\end{figure}
For example, the circuit in Fig.~\ref{fig:example-cnot} can be expressed as
\begin{verbatim}
f = cnot(x(I,0),0,1)
\end{verbatim}
Lambda calculus allows representing and working with these kinds of expressions efficiently, so that each function and its arguments are associated to the leaves of a tree and new programs can be obtained by modifying existing trees.
In this language, the previous function is expressed as
\begin{verbatim}
(lambda (lambda (lambda (cnot (x $0 $1) $1 $2))))
\end{verbatim}
where \$0 is the initial input circuit and \$i the qubit index (increased by one).
For the technical description of the lambda calculus programs as trees and their advantages, we refer to the DreamCoder paper~\cite{ellis_dreamcoder_2021}.

\section{Probabilistic framing}
\label{app:probabilistic-framing}
In this section, we give some more details about the probabilistic framing that yields the optimization objective of our algorithm, while still referring to ~\cite{ellis_dreamcoder_2021} for the full treatment.
We want to find the optimal set of gates $G^*$ and the optimal single gate probabilities $\theta^* = \{\theta_g, \forall g \in G\}$.
For this purpose, we perform Bayesian inference using Bayes theorem~\cite{papoulis_probability_2009}.
The posterior reads
\begin{equation}
    \label{eq:app-posterior}
    P(G,\theta | U)  = \frac{P(U|G,\theta) P(G,\theta)}{P(U)},
\end{equation}
where $P(G,\theta)$ is the prior, $P(U|G,\theta)$ is the likelihood, and $P(U)$ the marginal probability of the evidence.
We can consider $U$ as constant because it does not depend on this set of gates, hence we drop the denominator from the above equation (Eq.~\ref{eq:optimization-goal}):
\begin{equation}
    (G^* , \theta^*) = \arg \max_{G,\theta} P(G, \theta |U).
\end{equation}
By definition, we can factorize the joint distribution as $P(G,\theta)=P(G)P(\theta|G)$, and write the likelihood as
\begin{equation}
    P(U|G,\theta)=\prod_u P(u|G,\theta),
\end{equation}
where $P(u|G,\theta)$ is the probability of a specific unitary $u$ given $G$ and $\theta$, because of the independence between different unitary matrices in the dataset.
The probability of a specific unitary can in turn be written as a function of the probability of generating it from sequences of gates in $G$:
\begin{equation}
    P(u|G,\theta) = \sum_c P(u|c)P(c|G,\theta),
\end{equation}
where the sum over $c$ is intended over all the circuits that can be generated by assembling sequences of gates in $G$.
In our case, each unitary is deterministically induced by a circuit, thus
\begin{equation}
    P(u|c) = \mathbb{1}[\mathcal{U}(c)=u].
\end{equation}
The probability of a circuit given a gate set can be rewritten as a function of the gate probabilities $\theta$.
Indeed, a circuit is a sequence of gates, each applied to a set of qubits.
In general, the probability of the $i^{th}$ gate could depend on all the previous gates, thus the probability of a sequence of gates would read $P(g_0)P(g_1|g_0)P(g_2|g_1,g_0)\ldots$.
In our case, we made the roughest approximation to consider these probabilities to be independent: $\prod_g\theta_g$.
The probability of a circuit is thus obtained by multiplying the gate sequence probability and the multiplicity factor that takes into account the number of inputs of the gate and the number of possible qubits to connect it to.
Therefore, we obtain Eq.~\ref{eq:circuit-probability}
\begin{align}
    P(c|G, \theta) & =\prod_{g\in c} \mathbbb{1}\left[ g\in G \right] \theta_g \chi(c,g).
\end{align}
If the resulting circuit is not valid, for example, if the inputs of a CNOT gate are repeated, it is automatically discarded.
In this case, the $\chi$ factor would include an extra characteristic function that selects only the accepted circuits.
In particular, to obtain gate sequences of variable length, when sampling we add a terminating gate to $G$ with some probability $\theta_\text{end}$ and stop the sequence as soon as the gate gets extracted.

By putting it all together, we obtain the optimization objective in Eq.~\ref{eq:posterior}
\begin{align}
    P(G, \theta|U) \propto P(G)P(\theta|G)\prod_{u\in U}\sum_{c}\mathbbb{1}[\mathcal{U}(c)=u]P(c|G,\theta).
\end{align}

Our program synthesis algorithm systematically enumerates expression trees sequentially in decreasing probability order.
Therefore, instead of finding all the possible circuits that can produce a given unitary, we consider only the most likely $k$ (given the current set of gates $G$ and $\theta$).
We obtain the lower bound in Eq.~\ref{eq:lowerbound}
\begin{align}
     & P(G)
    P(\theta|G)\prod_{u\in U}\sum_{c\in \mathcal{B}_u}\mathbbb{1}[\mathcal{U}(c)=u]P(c|G,\theta)
\end{align}
just because we are neglecting smaller positive terms.
The maximization of this lower bound also implies the indirect maximization of the posterior.

The specific details of the enumeration algorithms are in the original DreamCoder paper \cite{ellis_dreamcoder_2021} and in the GitHub repository.
After we enumerated programs within a certain timeout, we start checking whether some of those programs can produce a circuit that is associated to one of the unitary matrices we want to decompose.
At each iteration, we consider a minibatch of $25$ unitaries from the whole training set and update the sets $\mathcal{B}_u$ for those unitaries.

\section{Implementation details}
The algorithm workflow is described in pseudocode in
Algorithm~\ref{alg:pseudocode}.
It takes as input the set of target unitary matrices to decompose, the initial set of elementary gates to use in the assembled circuits, the batch size of each iteration, the enumeration timeout in seconds, and the number of iterations.
The output is the final list of extracted gates.
To run the algorithm again and test the performance, it is enough to enumerate programs again for a given timeout and check against the test set.

\begin{algorithm}
    \caption{Unitary matrix decomposition algorithm}\label{alg:pseudocode}
    \KwData{\textit{target\_unitaries}, \textit{gate\_library}, \textit{batch\_size}, \textit{timeout}, \textit{N\_iterations}}
    \KwResult{\textit{gate\_library}}
    \For{$n<$\textit{N\_iterations}}{
        \textit{target\_unitaries\_batch} $\leftarrow$ get\_batch$($\textit{target\_unitaries}, \textit{batch\_size}$)$\;
        \textit{enumerated\_programs} $\leftarrow$ enumerate$($\textit{timeout}, \textit{gate\_library}$)$\;
        \textit{solutions} $\leftarrow$ check\_solutions$($\textit{enumerated\_programs, target\_unitary\_batch}$)$\;
        \textit{gate\_library} $\leftarrow$ update\_library$($\textit{elementary\_set, solutions}$)$\;
    }
\end{algorithm}
We ran the search in parallel on $32$ Xeon Gold 6130 CPUs.
Each of them explores disjoint subsets of the program space.
The enumeration runs for a fixed amount of time ($150s$).
The library building step can take a longer amount of time, according to the number of found solutions and the number of tasks in the considered batch of tasks (in our case, $25$).
This phase can last up to about $15$ minutes and is not parallelized.
The effect of checking only against a few tasks at each iteration allows speeding up the library building phase since we limit the number of unitary matrices each enumerated circuit should be tested against.
Of course, an additional consequence is that more iterations are overall needed, since it takes $N/N_{\text{batch}}$ iterations on average just to check against all the tasks.
The algorithm learns to decompose new matrices, but it accounts for that only when those are selected as target tasks.
\begin{figure}
    \centering
    \includegraphics[height=3cm]{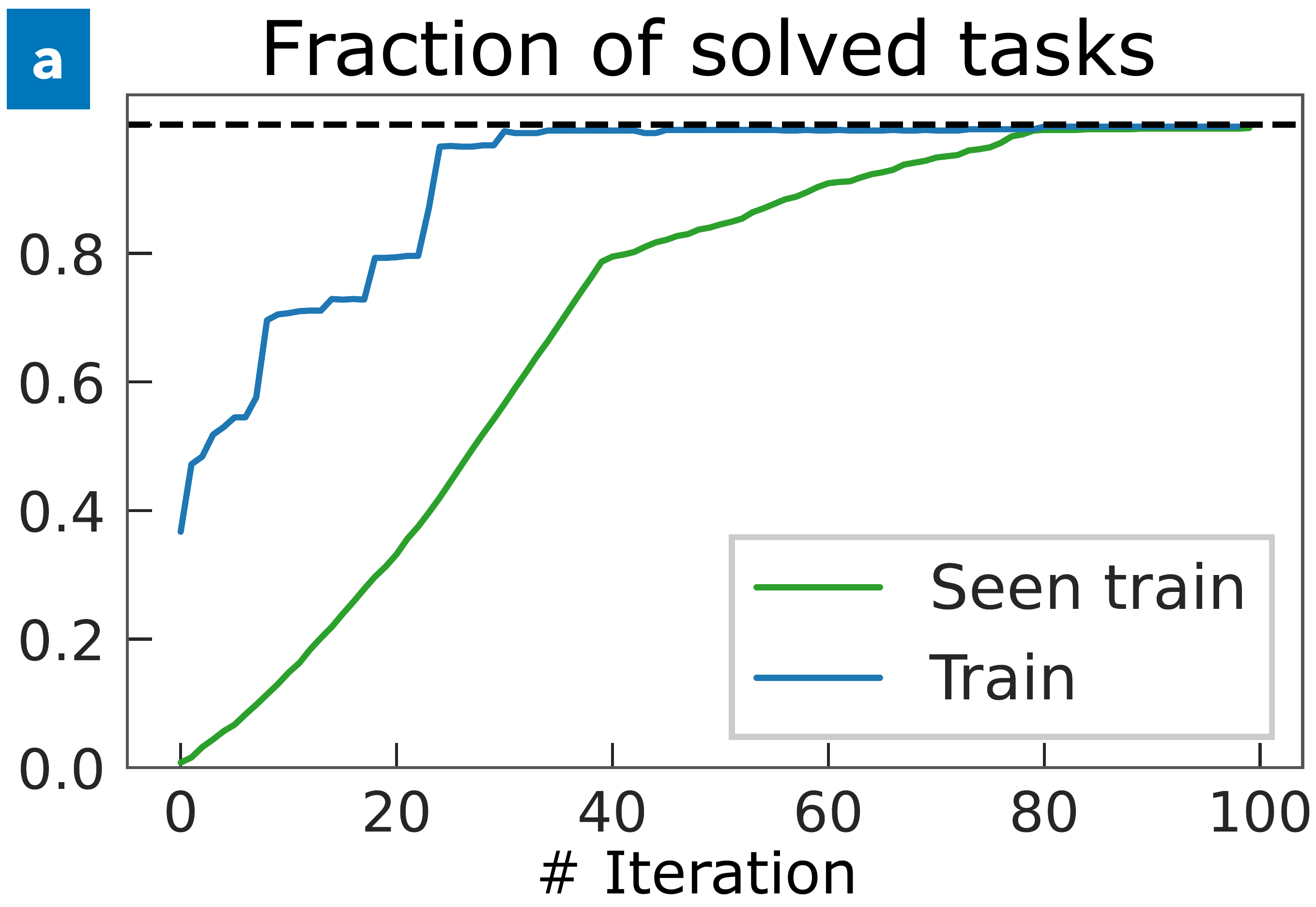}
    \caption{Estimated performance on the target set of matrices during algorithm iterations, calculated on the effectively seen matrices (in green), and on the complete set (in blue).
        Using batches of tasks adds some stochasticity to the algorithm (which helps to make it more robust) and speeds up the library building routine, but increases the number of required iterations. }
    \label{fig:appendix-fully-connected}
\end{figure}
As shown in Fig.~\ref{fig:appendix-fully-connected}, the performance that would be inferred at train time is, therefore, lower than the effective one on the train set, just because we don't check all the tasks at each iteration.
The "seen train" curve (calculated by considering only the decomposed matrices in the iteration batch) takes more time to take advantage of the discovered gates since at each iteration it is only evaluated on a subset of the total elements.
The same set, but completely evaluated at each iteration of the algorithm, is shown as "train" set (this is the same curve as in Fig.~\ref{fig:fully-connected}).
We see that, even if the algorithm can decompose all the matrices (for example after iteration $30$), it can still invent new gates to make the decomposition easier in terms of the number of used gates.

To produce plots of the circuits we use the qiskit library~\cite{at_av_qiskit_2021}.
The complete code to run the algorithm and reproduce the experiments is available on GitHub\footnote{\url{https://github.com/ellisk42/ec/tree/quantum_algorithms-simplified}}.

\section{Experiments}
In this Appendix, we show more details about the experiments we presented in the main text.
In Table~\ref{fig:fully-connected}, we show the list of all the gates that the algorithm extracted to solve the proposed tasks, in the case where no connectivity constraints were enforced (i.e. Figure~\ref{fig:fully-connected} in the main text).

\begin{longtblr}[
    caption = {List of extracted gates after $100$ iterations, with no connectivity constraints between qubits.},
    entry = {Extracted gates},
    label = {tblr:fully-connected},
    ]{
    colspec = {X[5mm]X[35mm]X[7cm]X}, hlines,
    rowhead = 1, rowfoot = 0,
    row{1} = {font=\bfseries},
    }

    \#
     & Gate representation
     & Expanded circuit
     & Program
    \\

    0
     & \includegraphics[height=15mm,valign=m]{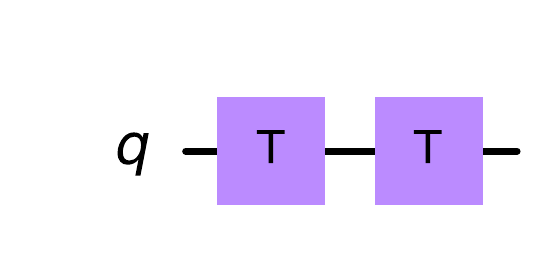}
     & \includegraphics[height=10mm,valign=m]{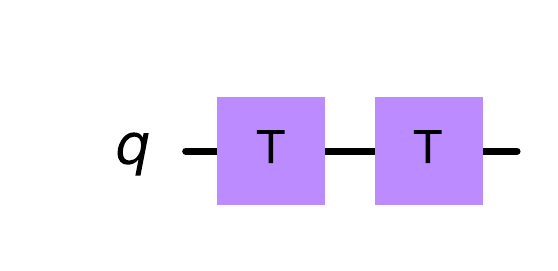}
     & \texttt{(lambda (lambda (t (t \$0 \$1) \$1)))}                                              \\

    1
     & \includegraphics[height=15mm,valign=m]{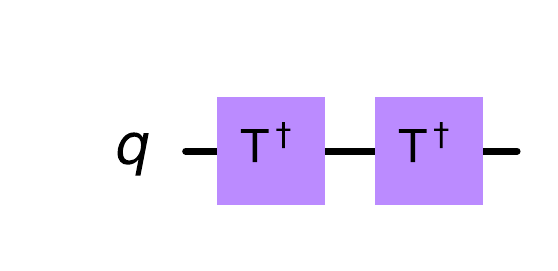}
     & \includegraphics[height=10mm,valign=m]{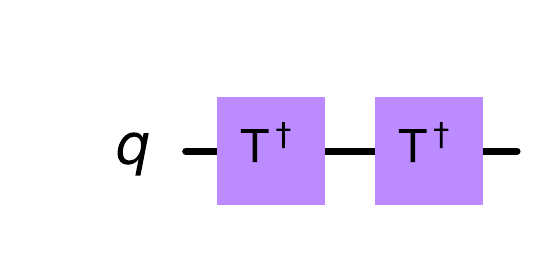}
     & \texttt{(lambda (lambda (tdg (tdg \$0 \$1) \$1)))}                                          \\

    2
     & \includegraphics[height=15mm,valign=m]{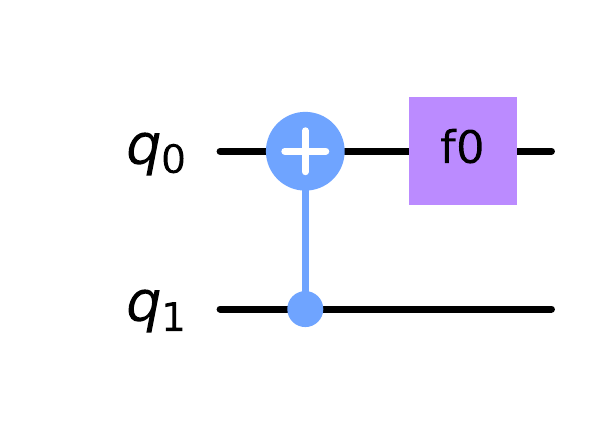}
     & \includegraphics[height=10mm,valign=m]{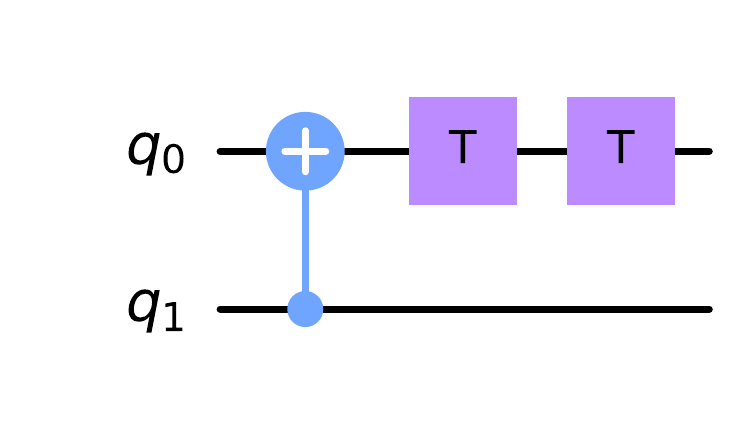}
     & \texttt{(lambda (lambda (lambda (f0 \$0 (cnot \$1 \$2 \$0)))))}                             \\

    3
     & \includegraphics[height=15mm,valign=m]{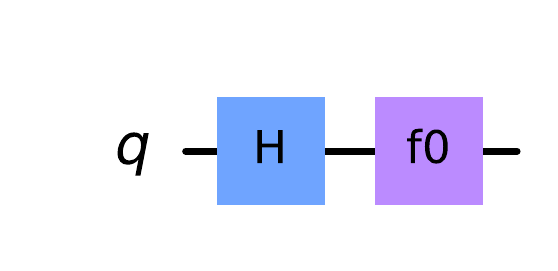}
     & \includegraphics[height=10mm,valign=m]{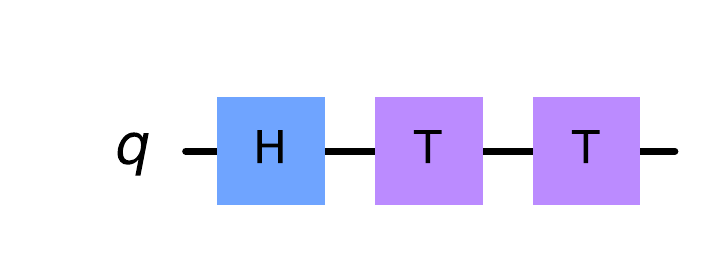}
     & \texttt{(lambda (lambda (f0 \$0 (h \$1 \$0))))}                                             \\

    4
     & \includegraphics[height=15mm,valign=m]{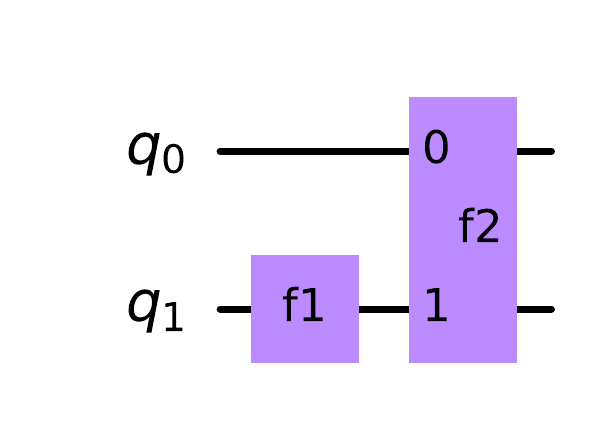}
     & \includegraphics[height=10mm,valign=m]{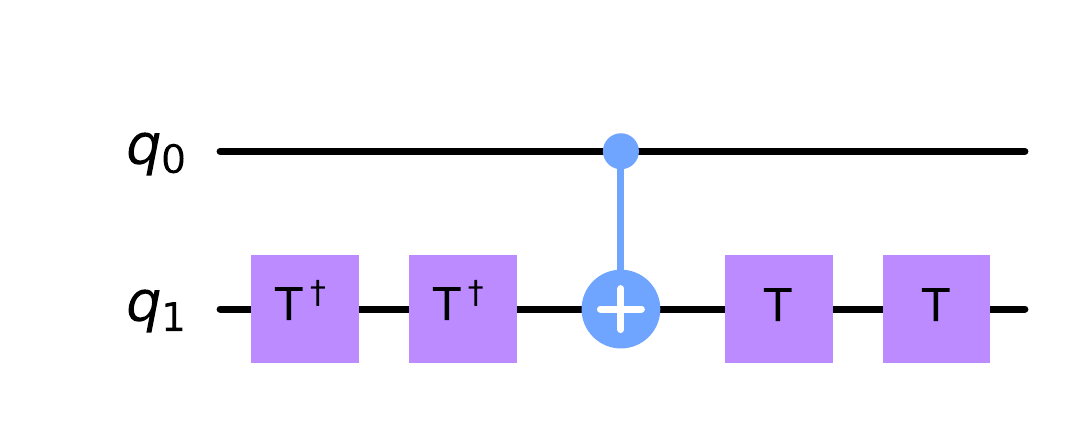}
     & \texttt{(lambda (lambda (lambda (f2 \$0 (f1 \$1 \$2) \$1))))}                               \\

    5
     & \includegraphics[height=15mm,valign=m]{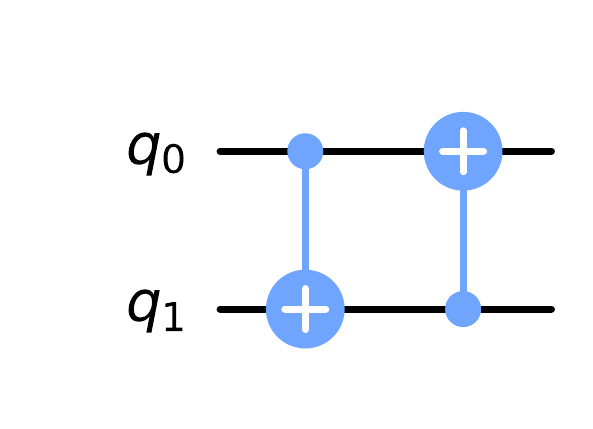}
     & \includegraphics[height=10mm,valign=m]{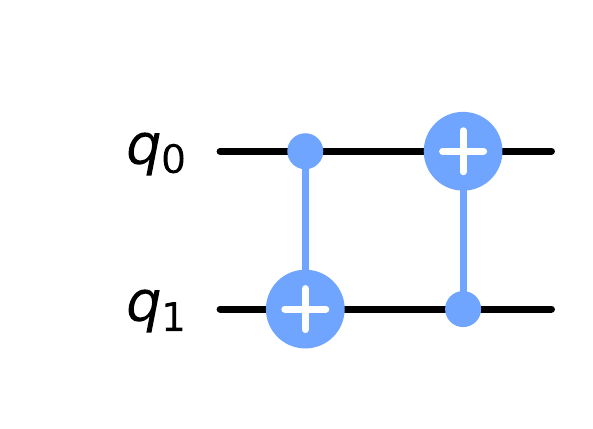}
     & \texttt{(lambda (lambda (lambda (cnot (cnot \$0 \$1 \$2) \$2 \$1))))}                       \\

    6
     & \includegraphics[height=15mm,valign=m]{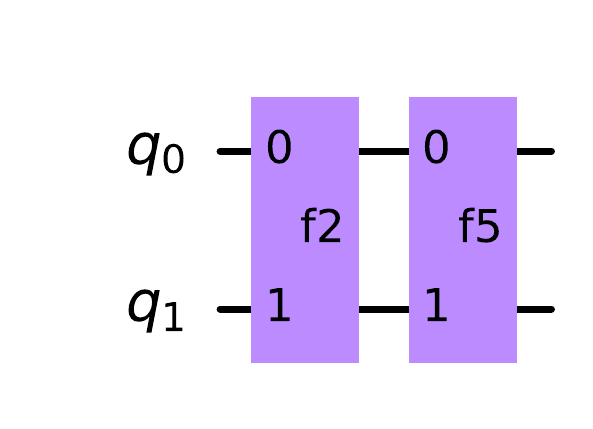}
     & \includegraphics[height=10mm,valign=m]{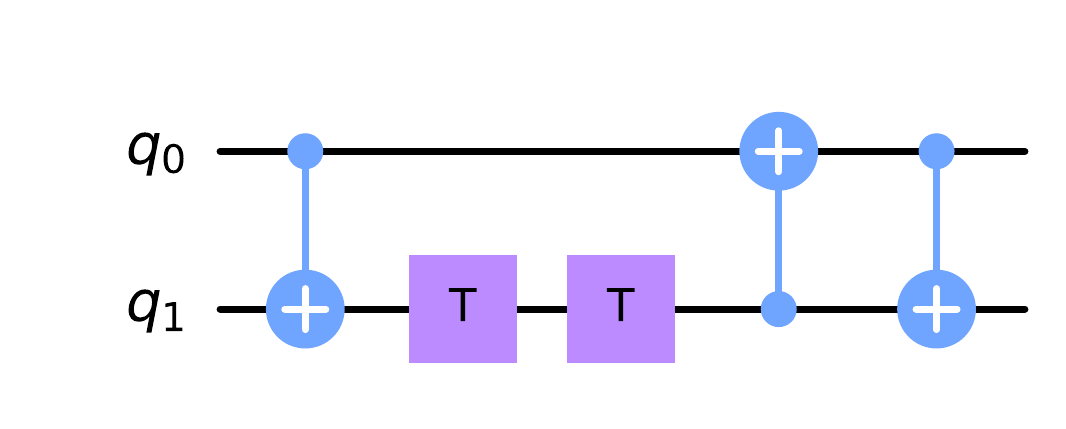}
     & \texttt{(lambda (lambda (lambda (f5 \$0 \$1 (f2 \$0 \$2 \$1)))))}                           \\

    7
     & \includegraphics[height=15mm,valign=m]{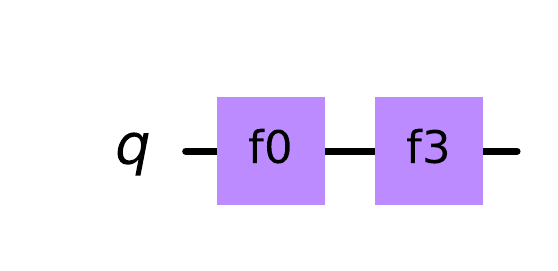}
     & \includegraphics[height=10mm,valign=m]{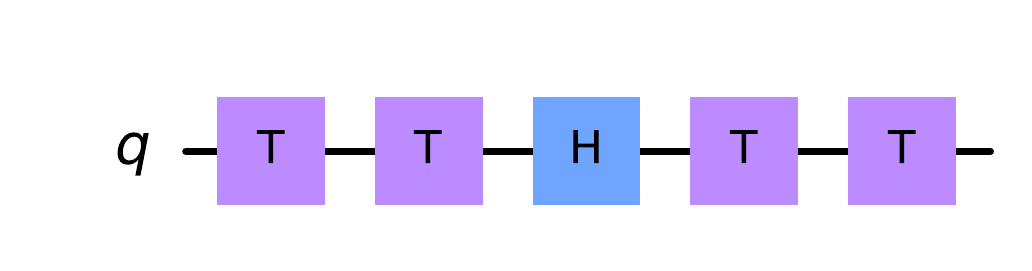}
     & \texttt{(lambda (lambda (f3 (f0 \$0 \$1) \$0)))}                                            \\

    8
     & \includegraphics[height=15mm,valign=m]{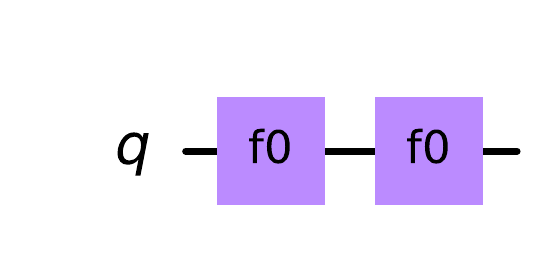}
     & \includegraphics[height=10mm,valign=m]{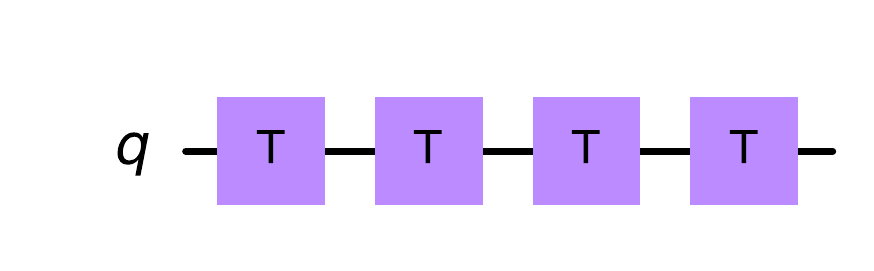}
     & \texttt{(lambda (lambda (f0 \$0 (f0 \$0 \$1))))}                                            \\

    9
     & \includegraphics[height=15mm,valign=m]{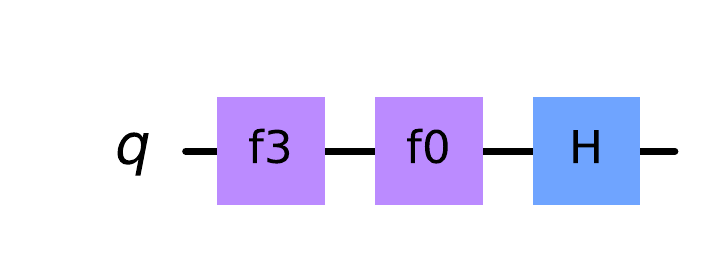}
     & \includegraphics[height=10mm,valign=m]{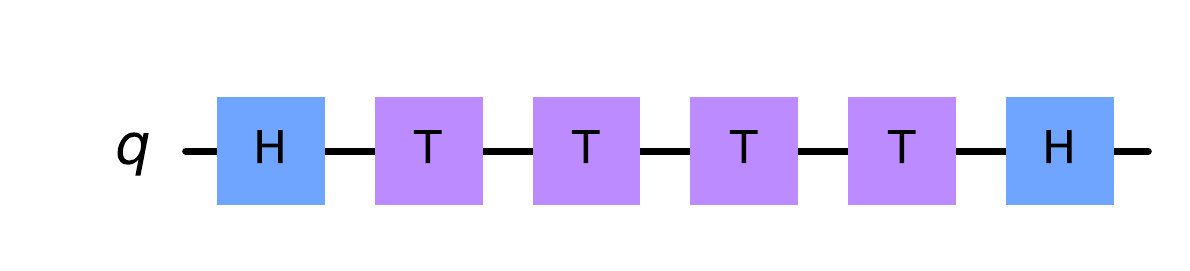}
     & \texttt{(lambda (lambda (h (f0 \$0 (f3 \$1 \$0)) \$0)))}                                    \\

    10
     & \includegraphics[height=15mm,valign=m]{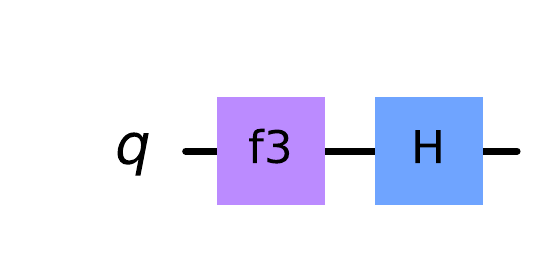}
     & \includegraphics[height=10mm,valign=m]{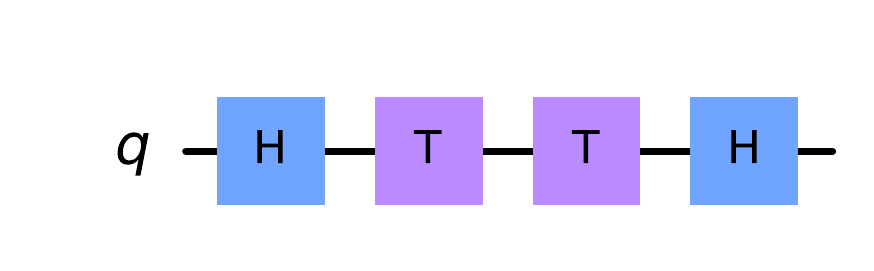}
     & \texttt{(lambda (lambda (h (f3 \$0 \$1) \$1)))}                                             \\

    11
     & \includegraphics[height=15mm,valign=m]{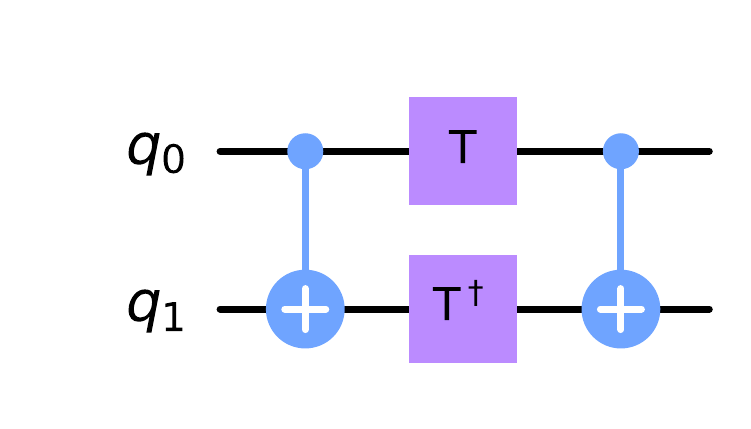}
     & \includegraphics[height=10mm,valign=m]{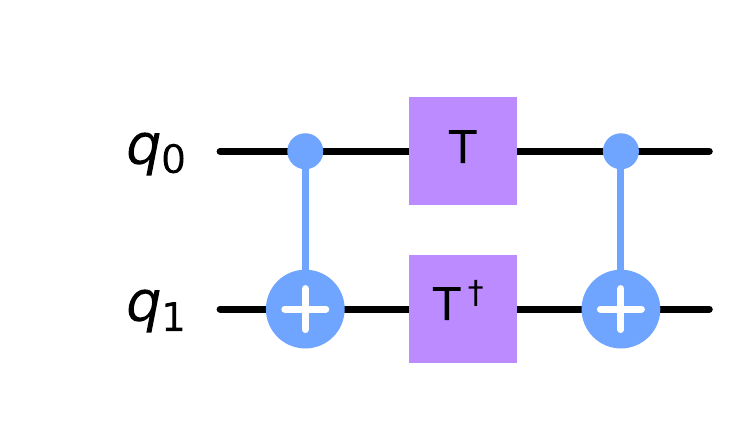}
     & \texttt{(lambda (lambda (lambda (cnot (t (tdg (cnot \$0 \$1 \$2) \$2) \$1) \$1 \$2))))}     \\

    12
     & \includegraphics[height=15mm,valign=m]{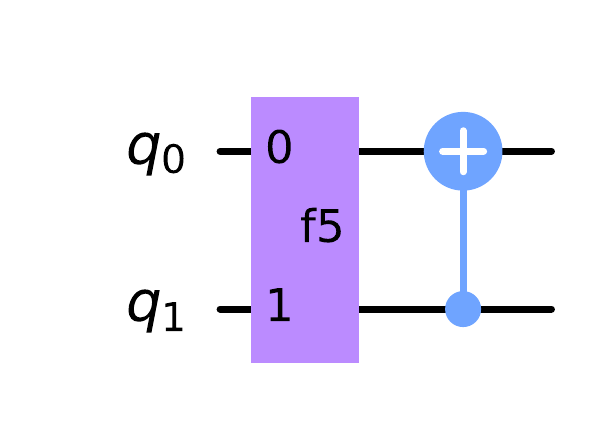}
     & \includegraphics[height=10mm,valign=m]{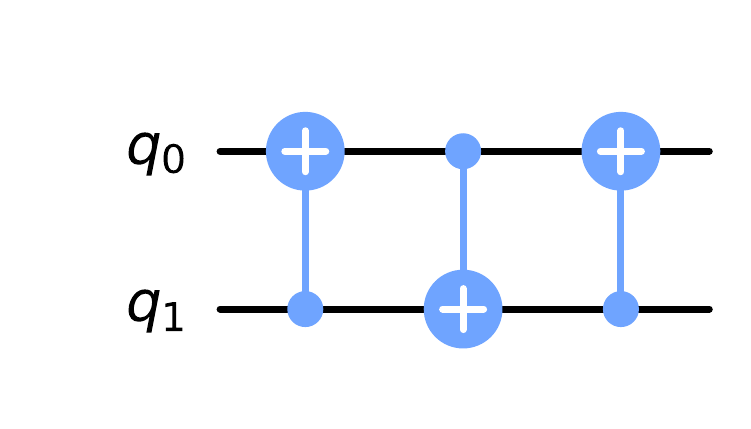}
     & \texttt{(lambda (lambda (lambda (cnot (f5 \$0 \$1 \$2) \$1 \$0))))}                         \\

    13
     & \includegraphics[height=15mm,valign=m]{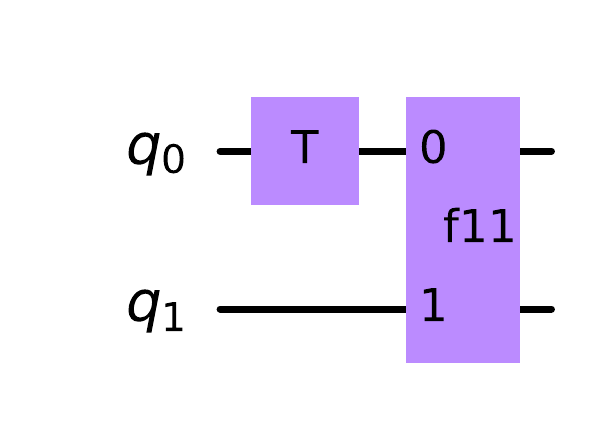}
     & \includegraphics[height=10mm,valign=m]{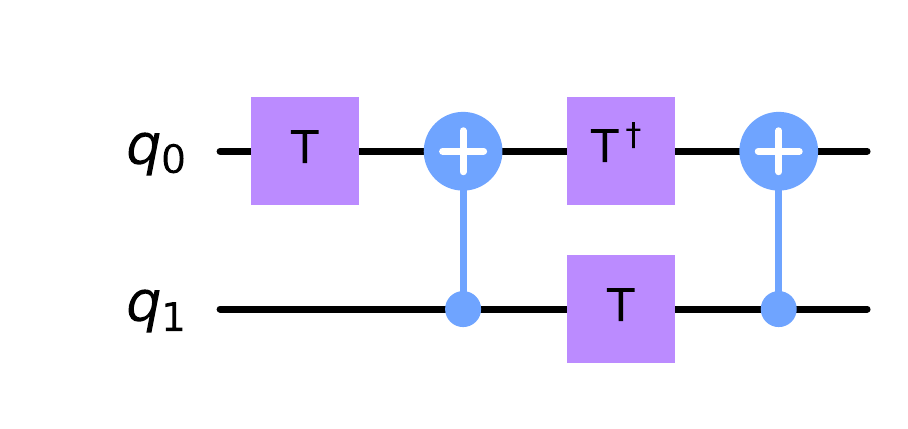}
     & \texttt{(lambda (lambda (lambda (f11 \$0 \$1 (t \$2 \$0)))))}                               \\

    14
     & \includegraphics[height=15mm,valign=m]{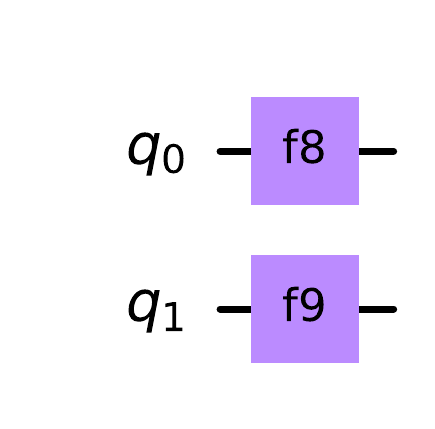}
     & \includegraphics[height=10mm,valign=m]{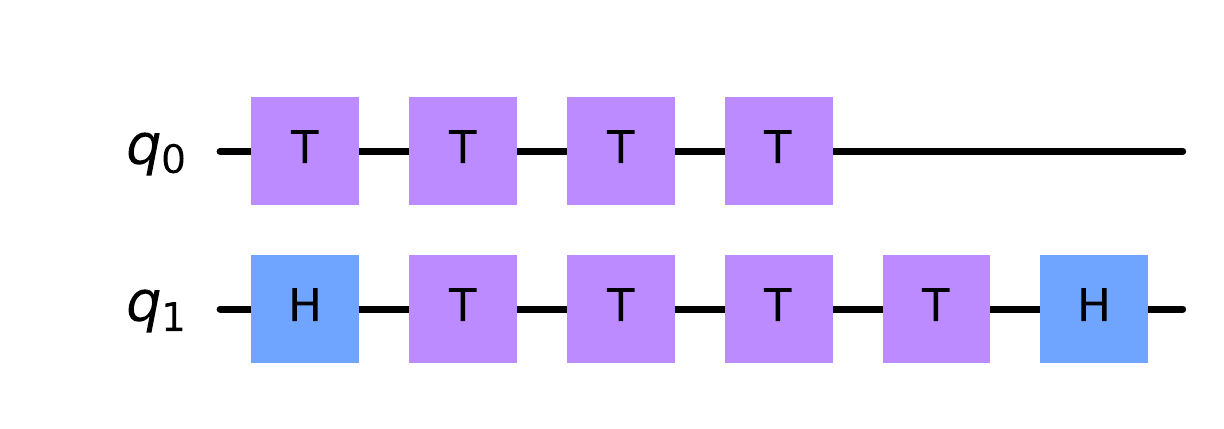}
     & \texttt{(lambda (lambda (f8 (f9 \$0 \$1))))}                                                \\

    15
     & \includegraphics[height=15mm,valign=m]{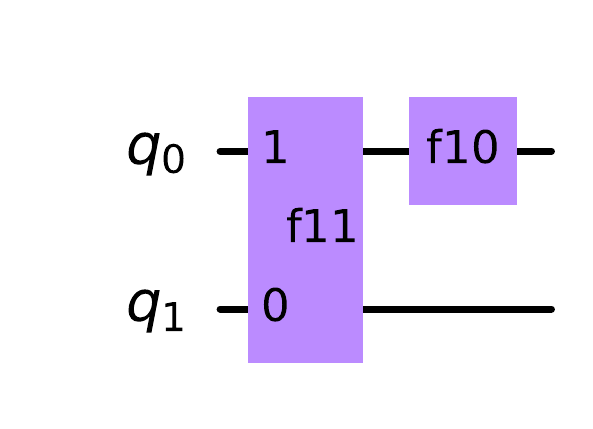}
     & \includegraphics[height=10mm,valign=m]{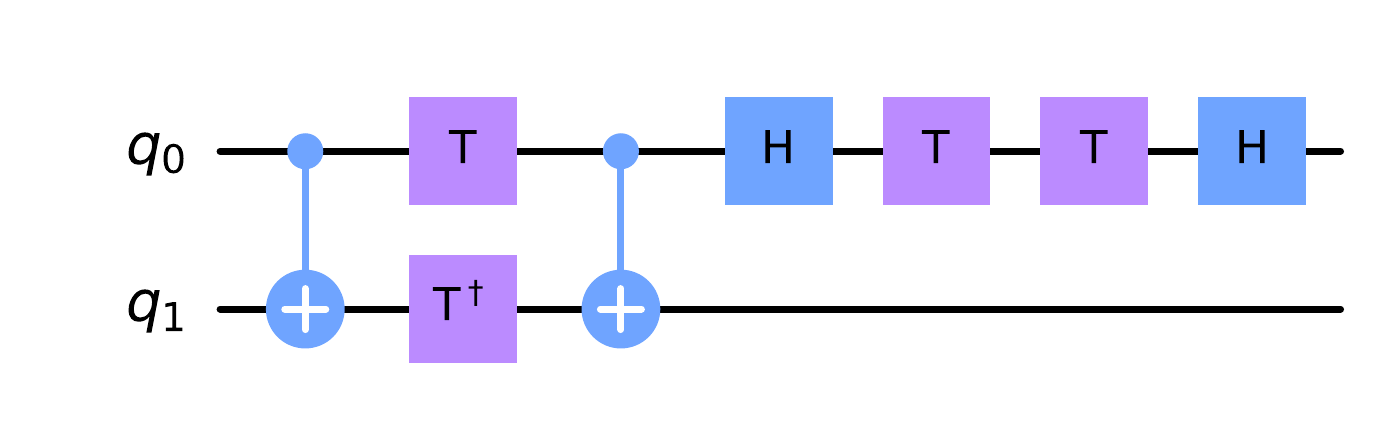}
     & \texttt{(lambda (lambda (lambda (f10 \$0 (f11 \$1 \$0 \$2)))))}                             \\

    16
     & \includegraphics[height=15mm,valign=m]{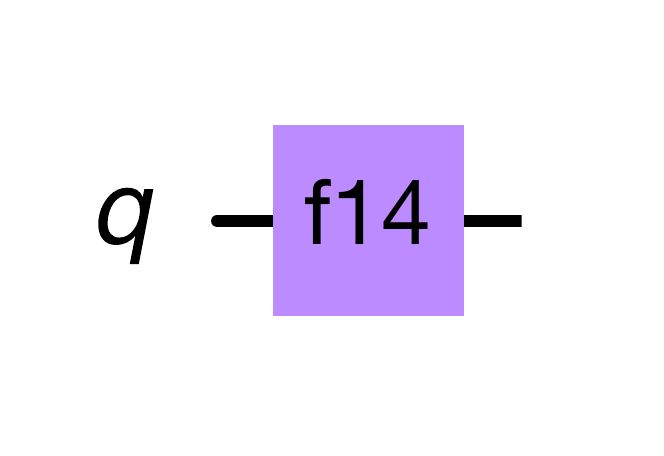}
     & \includegraphics[height=10mm,valign=m]{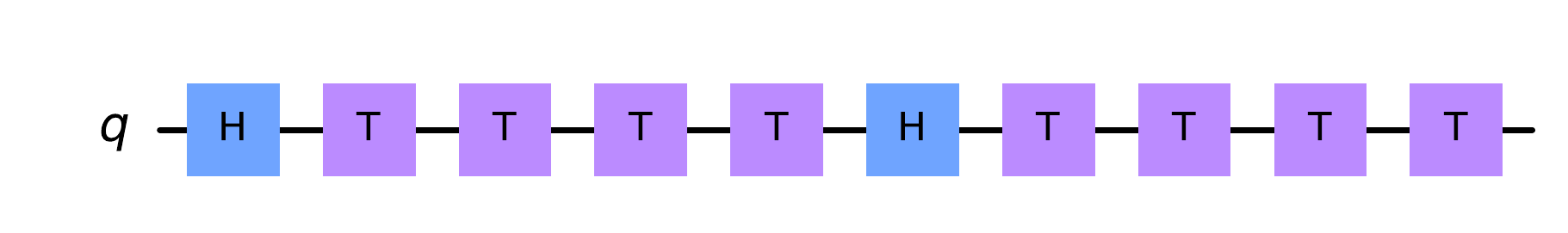}
     & \texttt{(lambda (lambda (f14 \$0 \$1 \$0)))}                                                \\

    17
     & \includegraphics[height=15mm,valign=m]{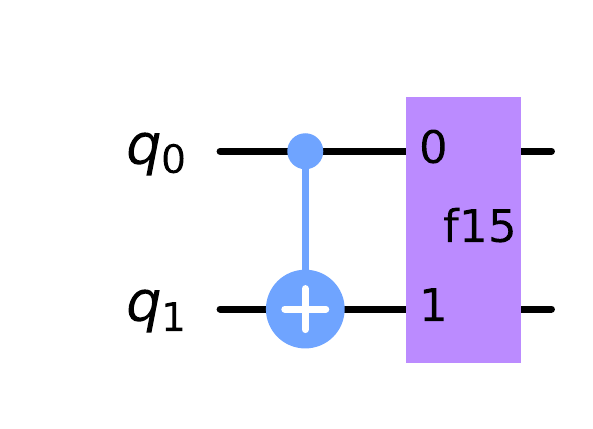}
     & \includegraphics[height=10mm,valign=m]{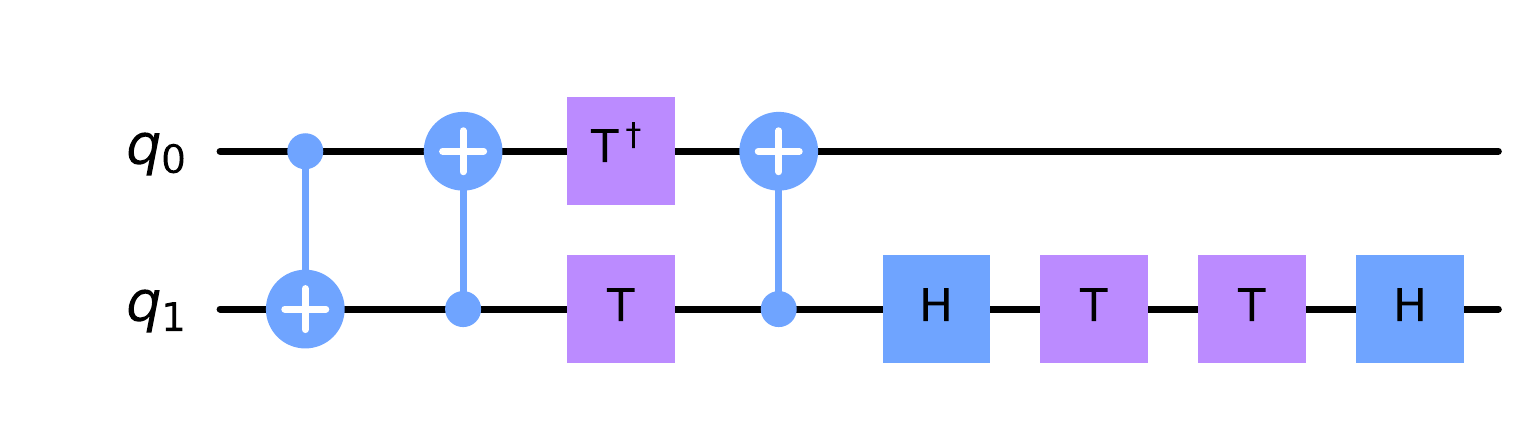}
     & \texttt{(lambda (lambda (lambda (f15 (cnot \$0 \$1 \$2) \$1 \$2))))}                        \\

    18
     & \includegraphics[height=15mm,valign=m]{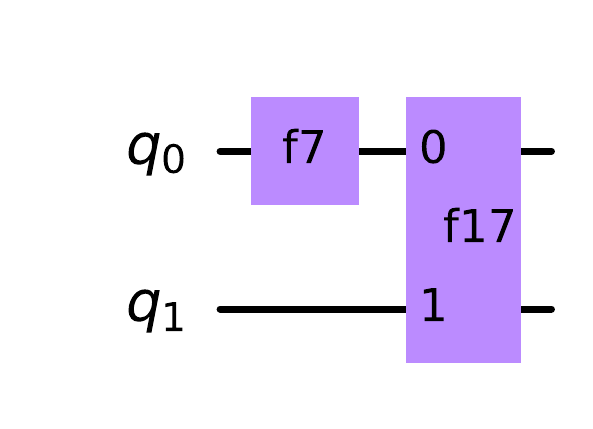}
     & \includegraphics[height=10mm,valign=m]{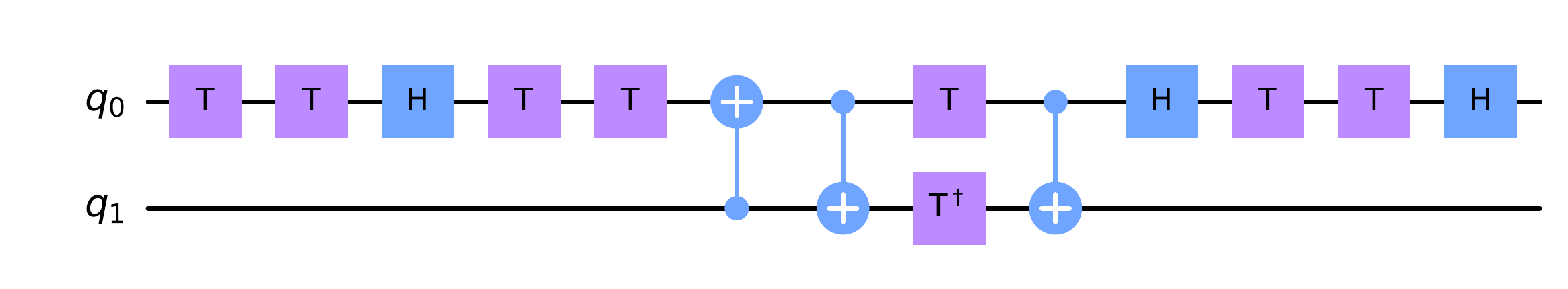}
     & \texttt{(lambda (lambda (lambda (f17 \$0 \$1 (f7 \$2 \$0)))))}                              \\

    19
     & \includegraphics[height=15mm,valign=m]{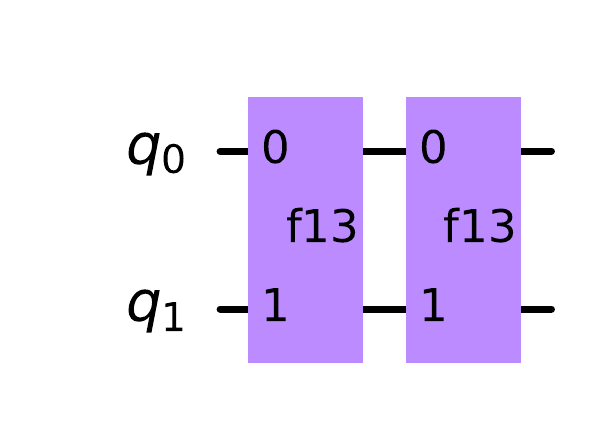}
     & \includegraphics[height=10mm,valign=m]{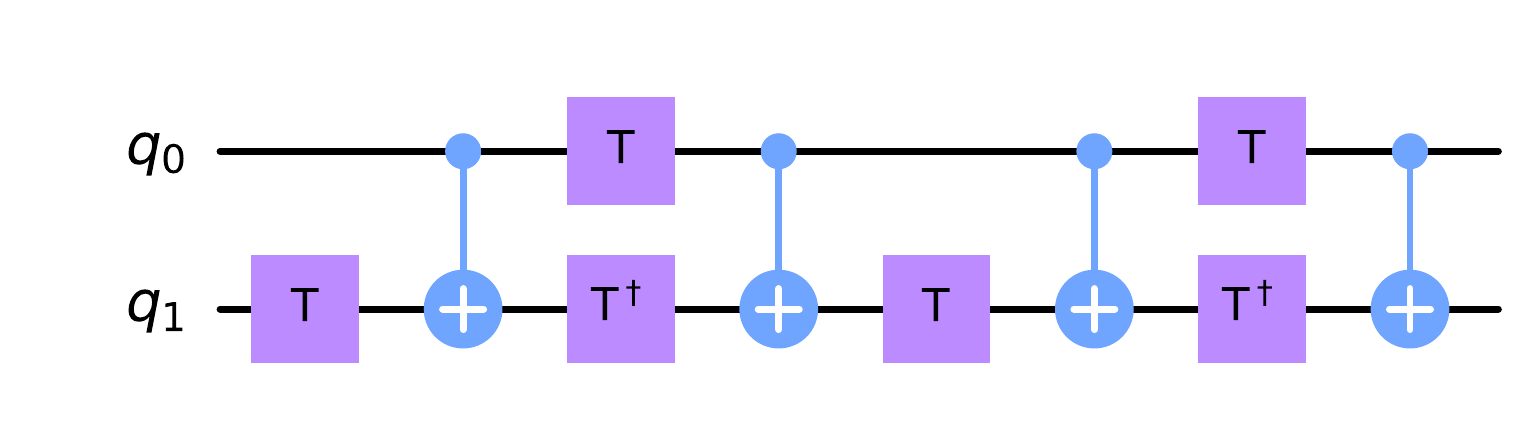}
     & \texttt{(lambda (lambda (lambda (f13 (f13 \$0 \$1 \$2) \$1 \$2))))}                         \\

    20
     & \includegraphics[height=15mm,valign=m]{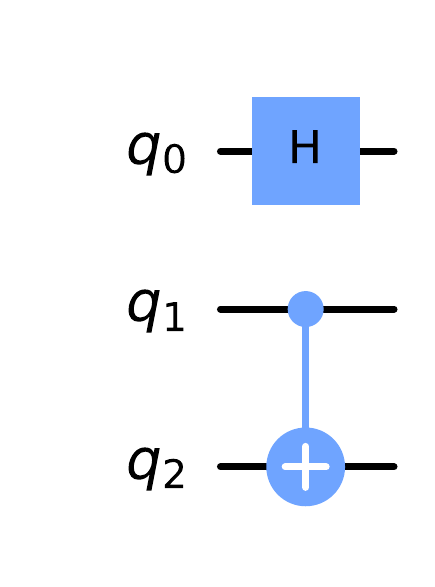}
     & \includegraphics[height=10mm,valign=m]{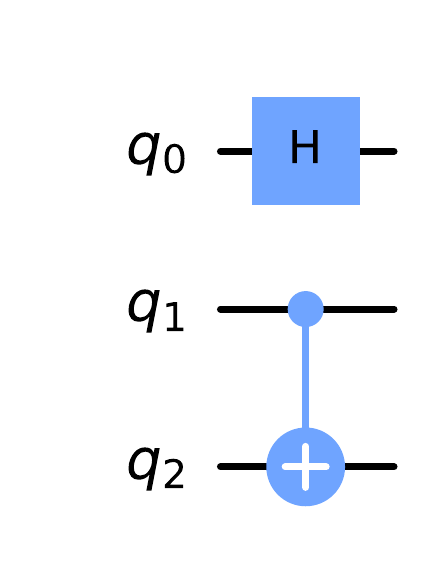}
     & \texttt{(lambda (lambda (lambda (h (cnot \$0 \$1 \$2)))))}                                  \\

    21
     & \includegraphics[height=15mm,valign=m]{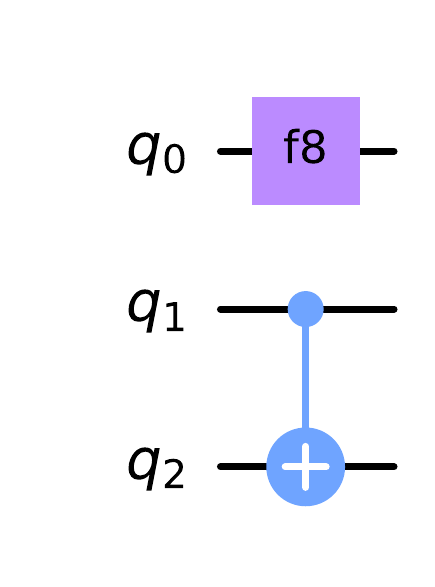}
     & \includegraphics[height=10mm,valign=m]{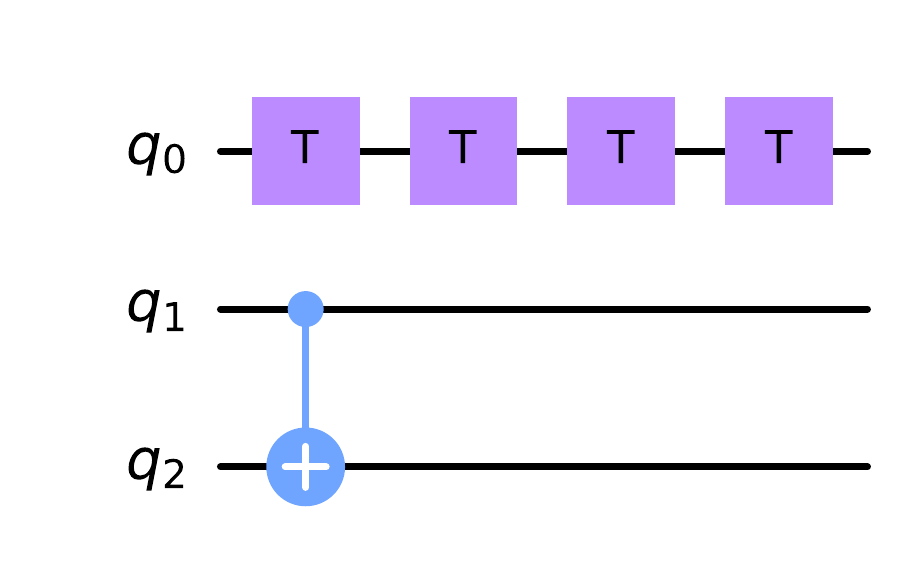}
     & \texttt{(lambda (lambda (lambda (f8 (cnot \$0 \$1 \$2)))))}                                 \\

    22
     & \includegraphics[height=15mm,valign=m]{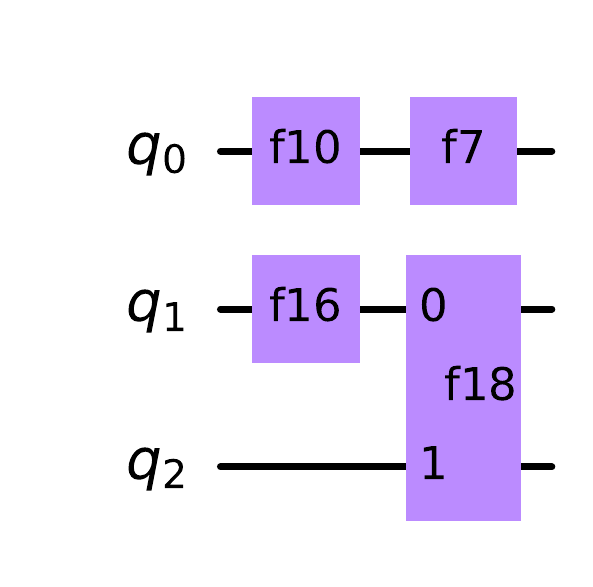}
     & \includegraphics[height=10mm,valign=m]{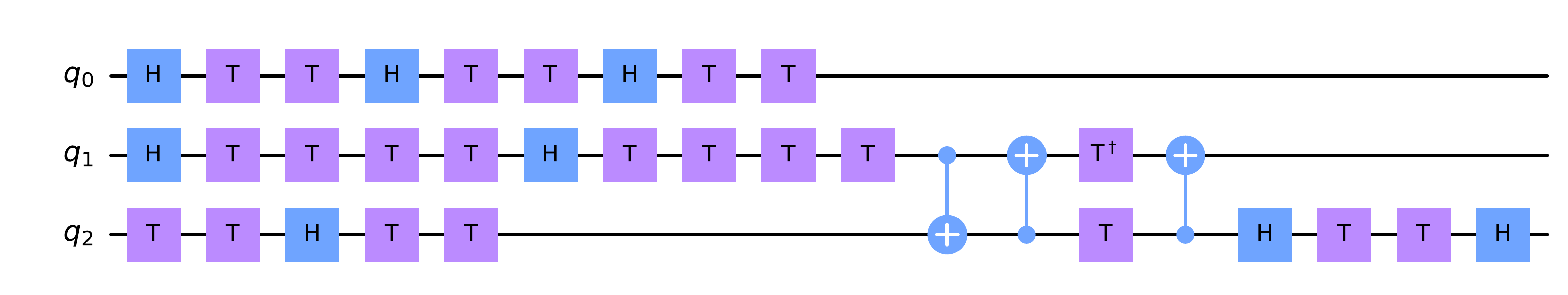}
     & \texttt{(lambda (lambda (lambda (lambda (f7 (f10 \$0 (f18 (f16 \$1 \$2) \$2 \$3)) \$0)))))} \\
\end{longtblr}

Table~\ref{tblr:fully-connected-tasks} shows some examples of unitary matrices in our set (expressed in terms of the circuit that we used to generate the matrix) and the decompositions proposed by our algorithm.
We see that the algorithm also finds unexpected ways to decompose the associated unitary matrix, which does not necessarily involve the use of exactly the same blocks we used to build it.

\begin{longtblr}[
    caption = {Examples of decomposed matrices.},
    entry = {Decomposed matrices},
    label = {tblr:fully-connected-tasks},
    ]{
    colspec = {X[15mm]X[25mm]X}, hlines,
    rowhead = 1, rowfoot = 0,
    row{1} = {font=\bfseries},
    }

    Target
     & Decomposition
     & Expanded decomposition
    \\

    \includegraphics[height=15mm,valign=m]{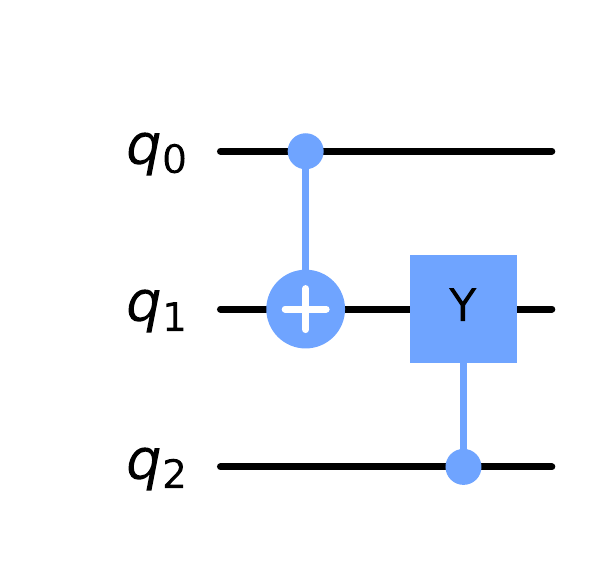}
     & \includegraphics[height=15mm,valign=m]{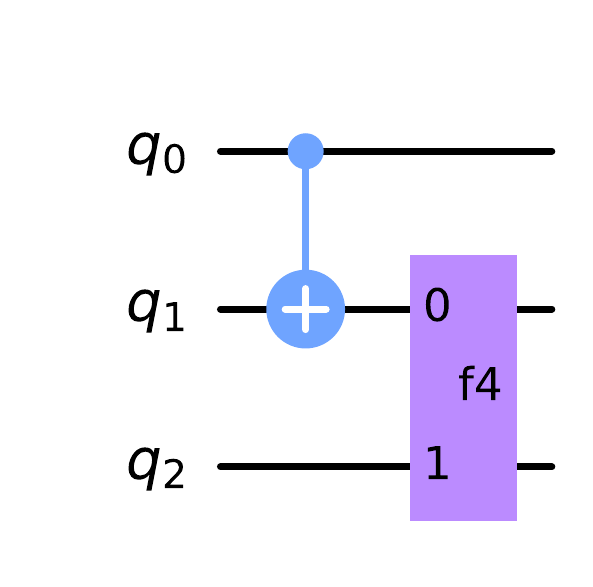}
     & \includegraphics[height=15mm,valign=m]{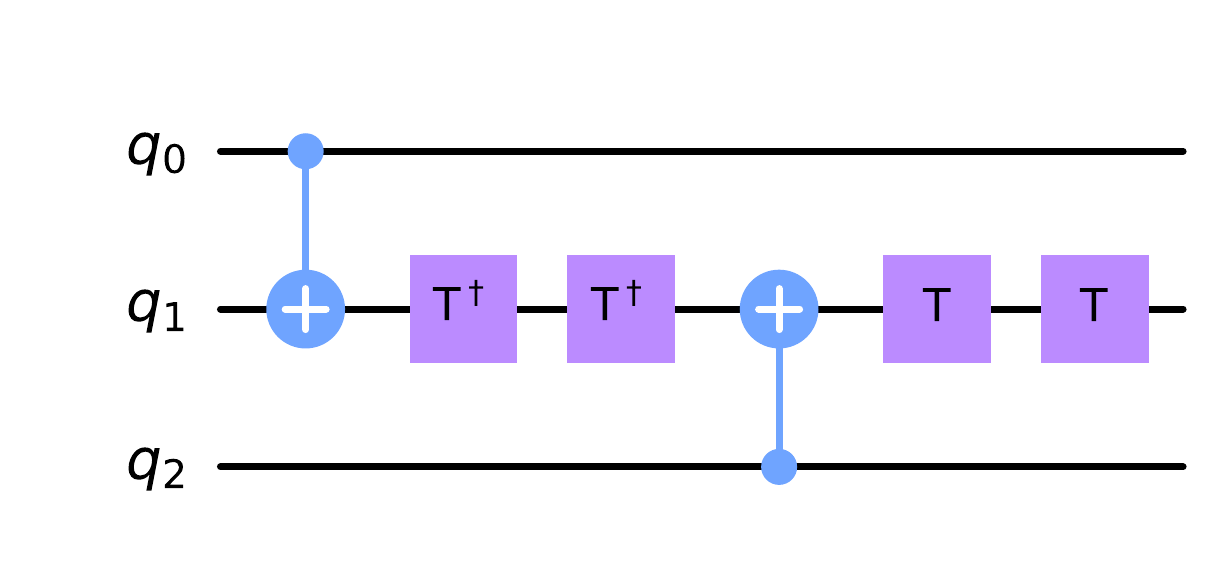}           \\

    \includegraphics[height=15mm,valign=m]{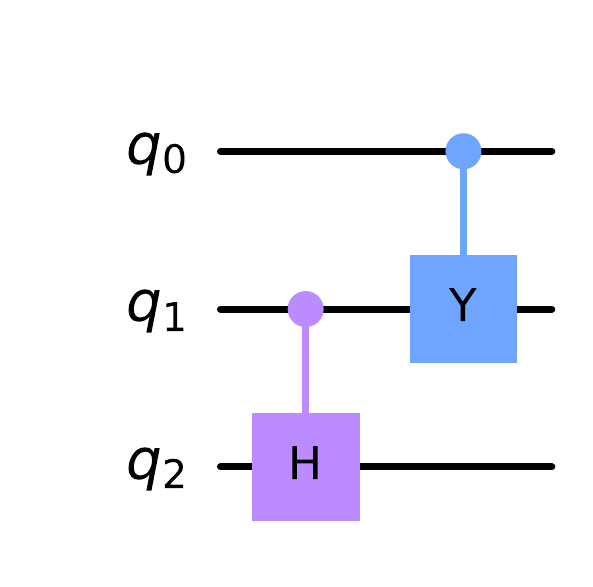}
     & \includegraphics[height=15mm,valign=m]{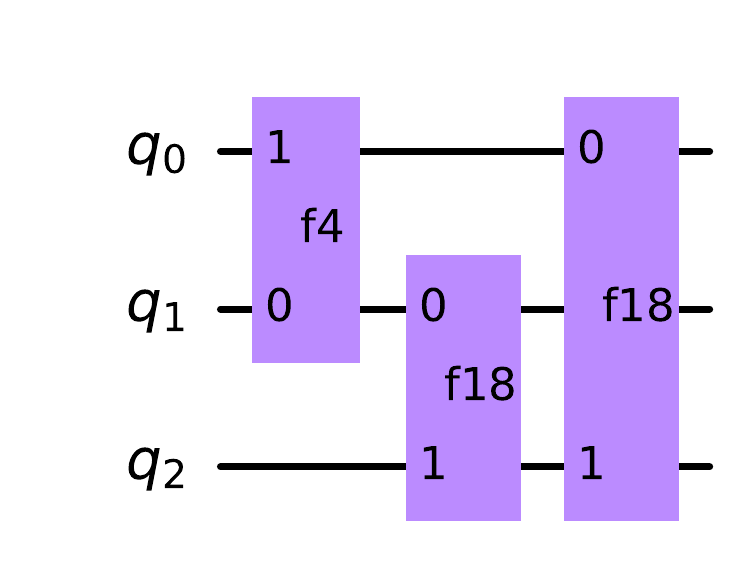}
     & \includegraphics[height=15mm,valign=m]{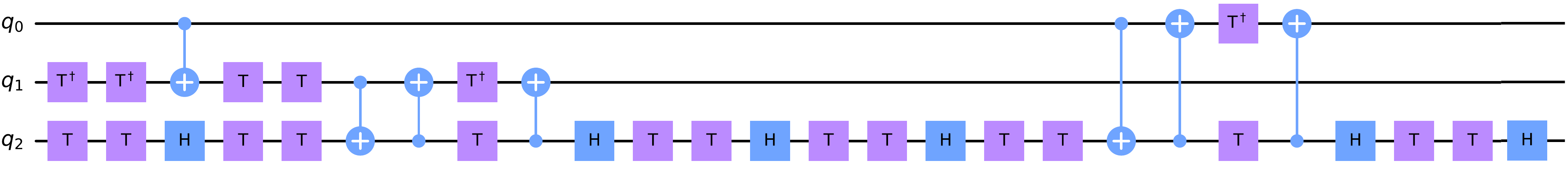}           \\

    \includegraphics[height=15mm,valign=m]{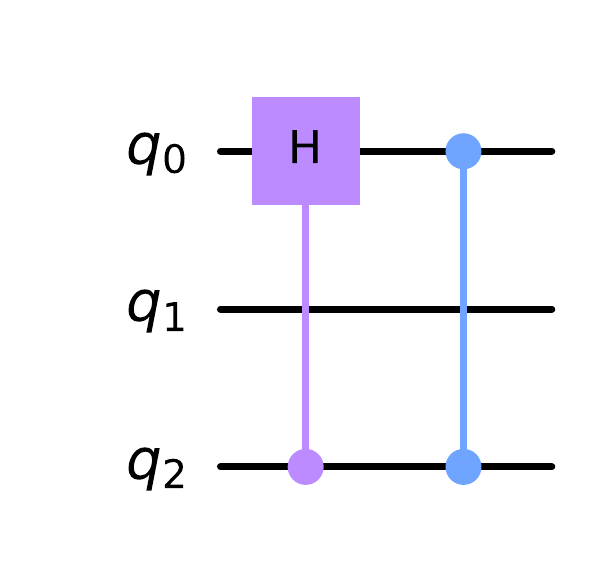}
     & \includegraphics[height=15mm,valign=m]{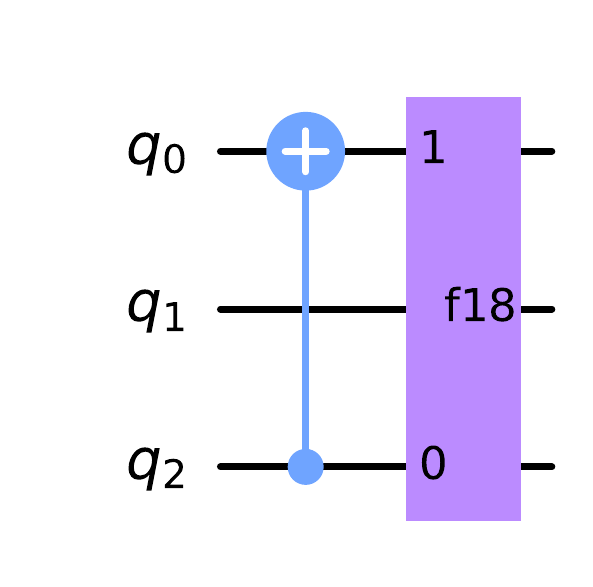}
     & \includegraphics[height=15mm,valign=m]{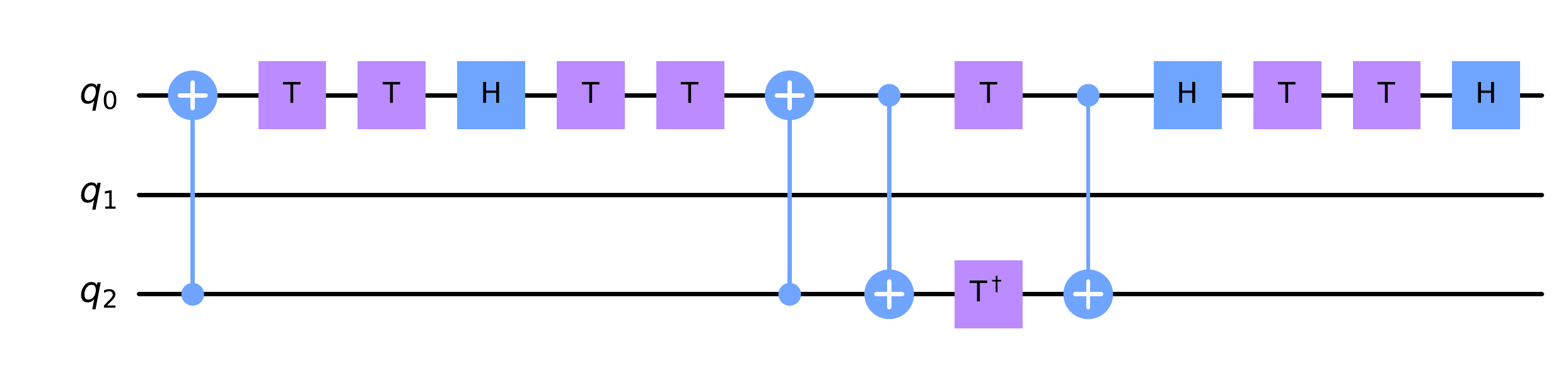}           \\

    \includegraphics[height=15mm,valign=m]{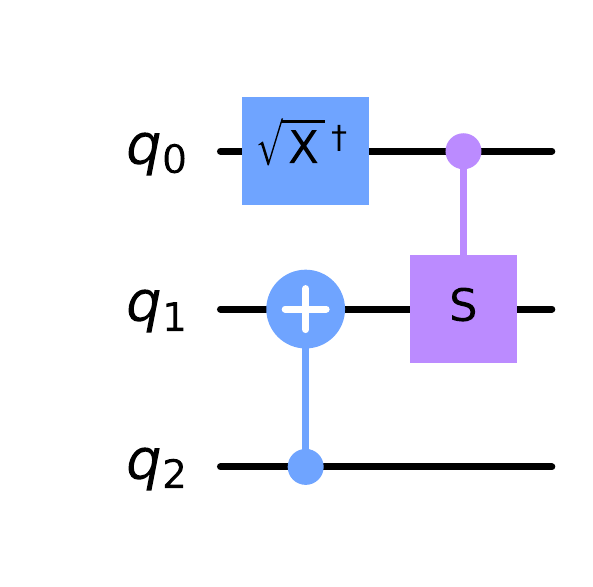}
     & \includegraphics[height=15mm,valign=m]{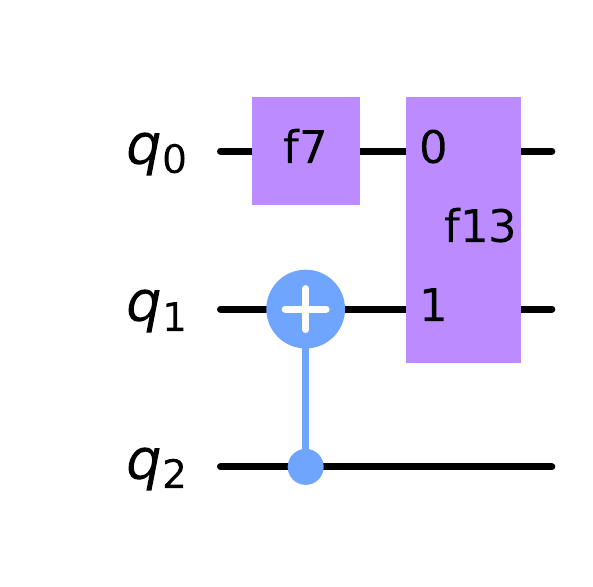}
     & \includegraphics[height=15mm,valign=m]{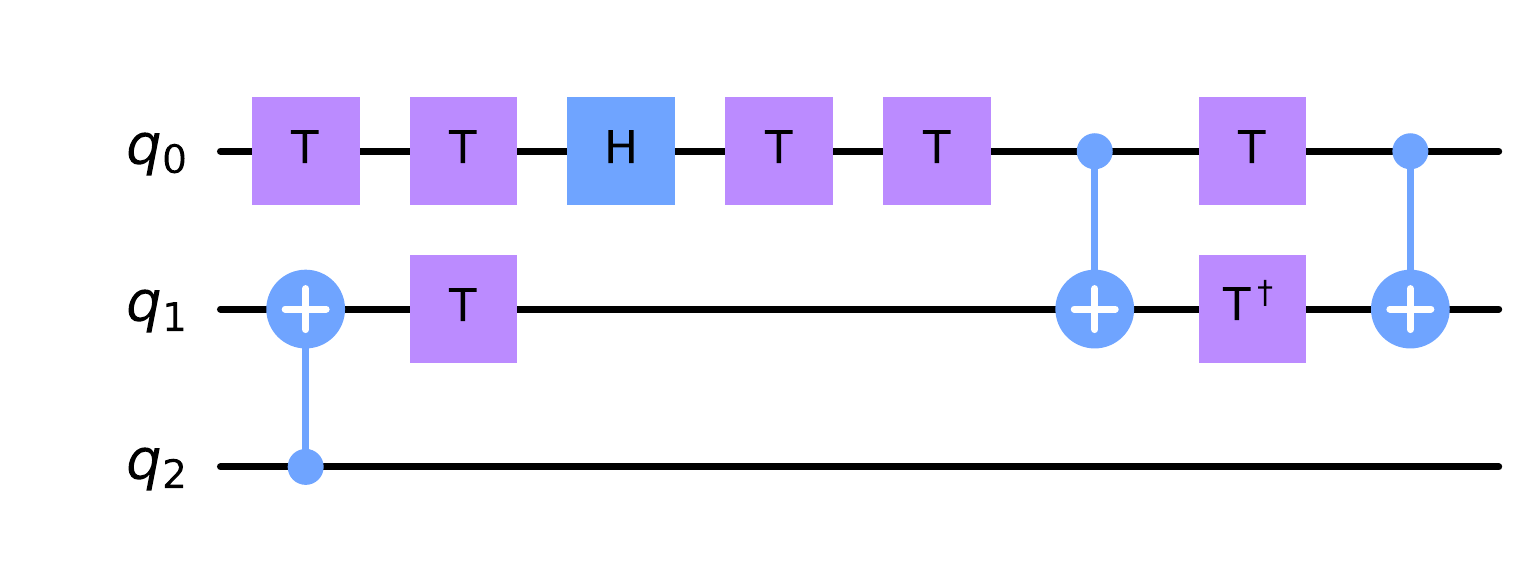}          \\

    \includegraphics[height=15mm,valign=m]{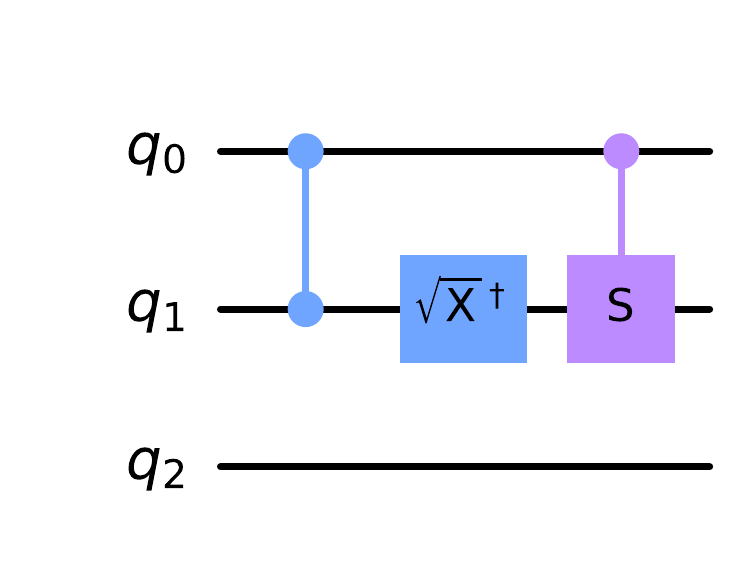}
     & \includegraphics[height=15mm,valign=m]{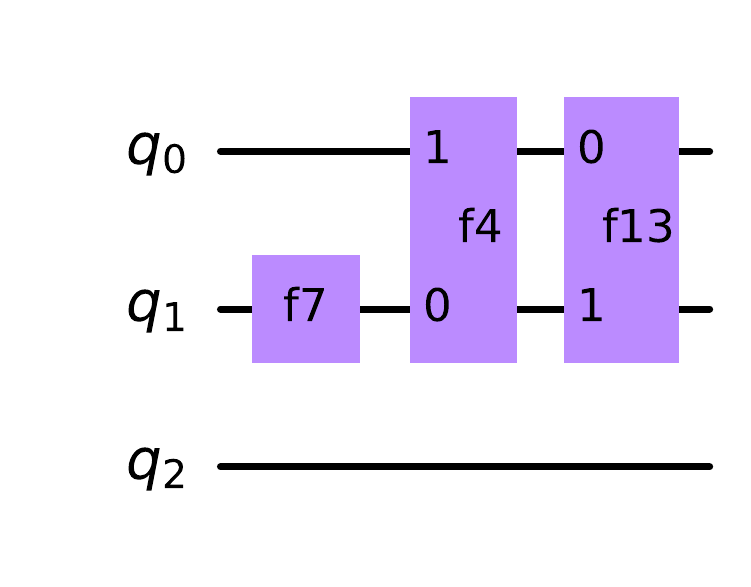}
     & \includegraphics[height=15mm,valign=m]{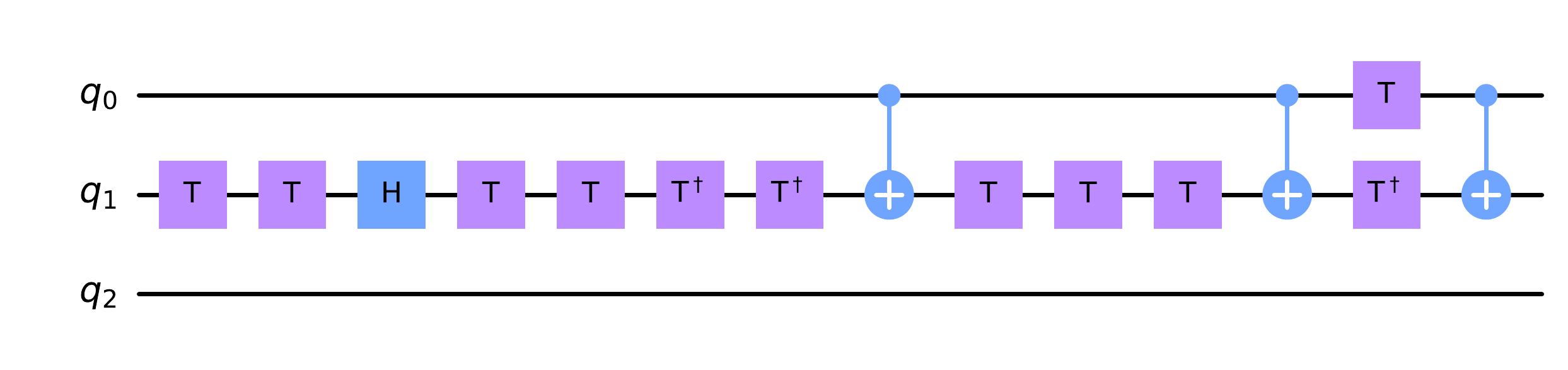}          \\

    \includegraphics[height=15mm,valign=m]{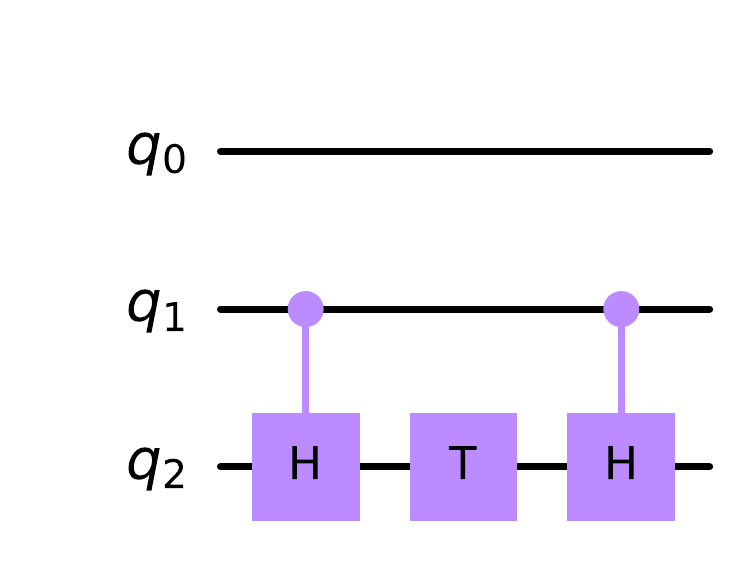}
     & \includegraphics[height=15mm,valign=m]{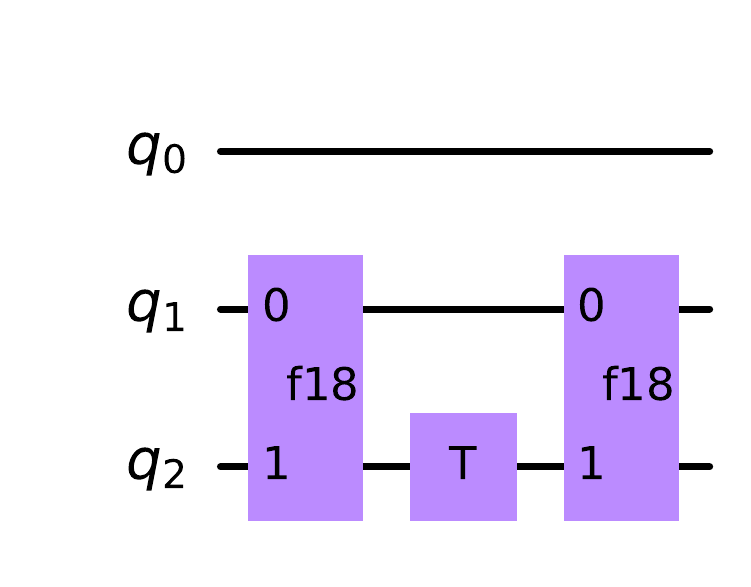}
     & \includegraphics[height=10mm,valign=m]{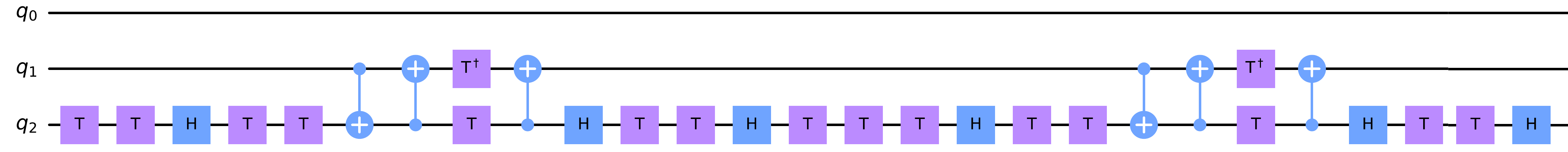}          \\

    \includegraphics[height=15mm,valign=m]{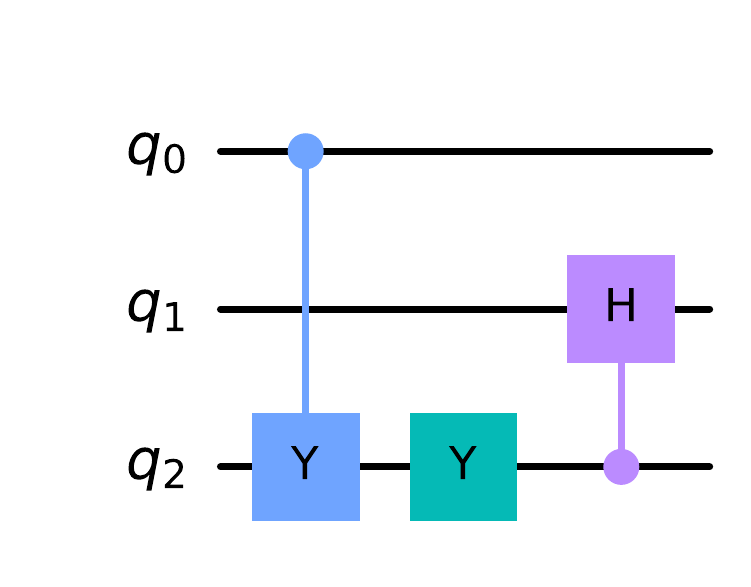}
     & \includegraphics[height=15mm,valign=m]{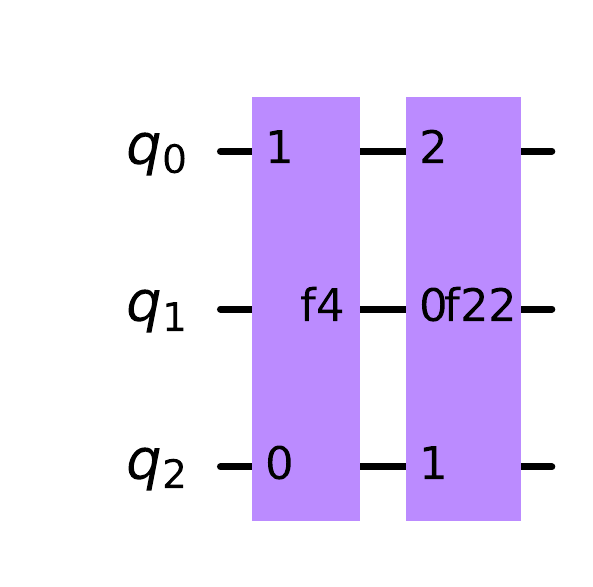}
     & \includegraphics[height=15mm,valign=m]{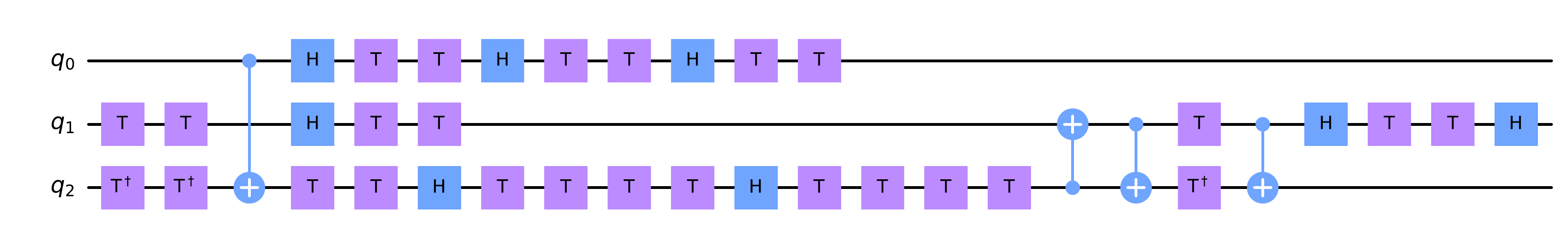}          \\

    \includegraphics[height=15mm,valign=m]{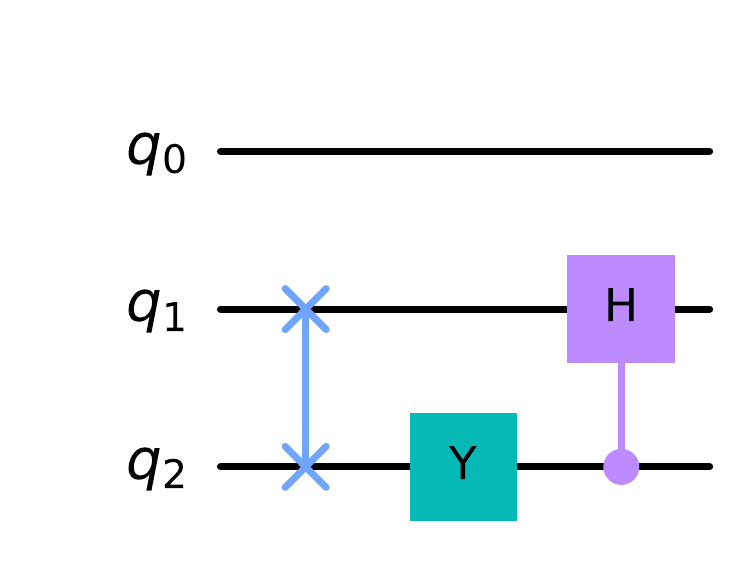}
     & \includegraphics[height=15mm,valign=m]{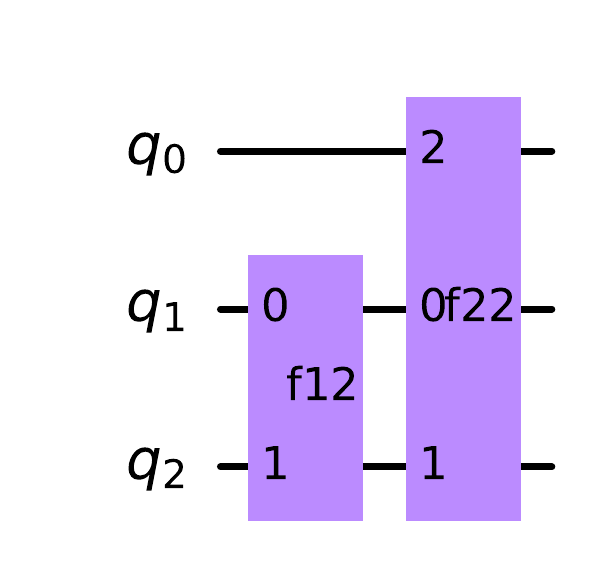}
     & \includegraphics[height=15mm,valign=m]{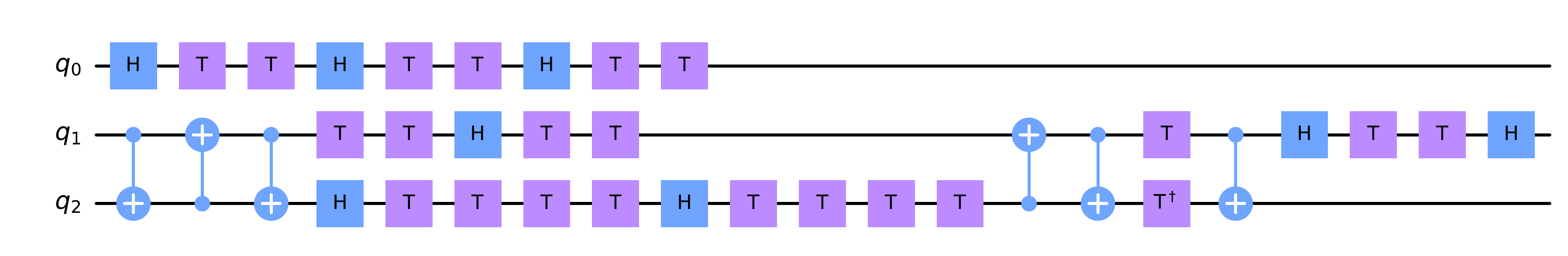}          \\

    \includegraphics[height=15mm,valign=m]{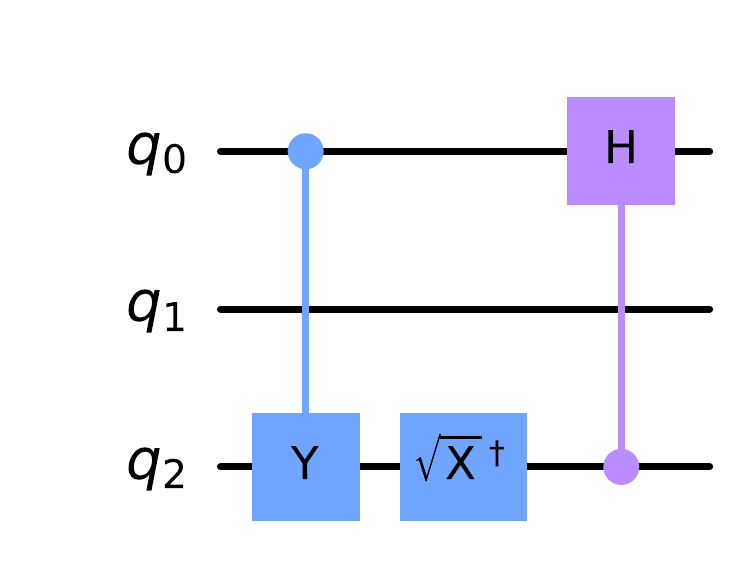}
     & \includegraphics[height=15mm,valign=m]{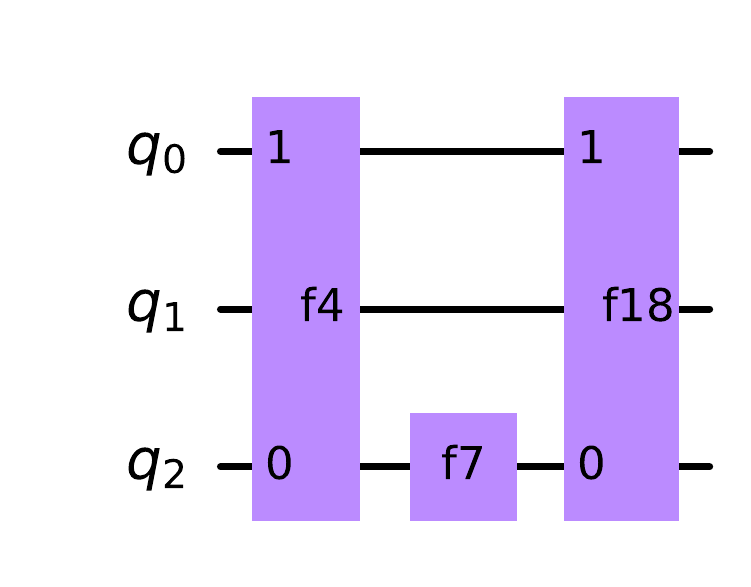}
     & \includegraphics[height=15mm,valign=m]{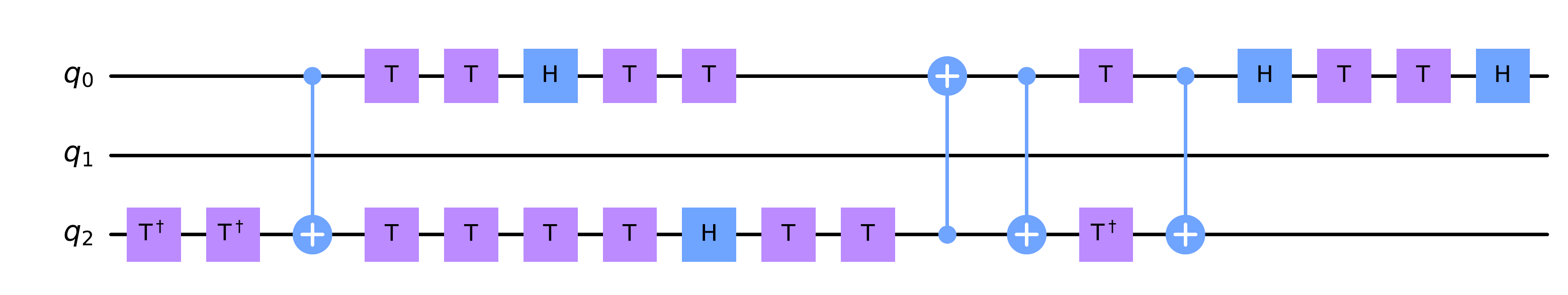}          \\

    \includegraphics[height=15mm,valign=m]{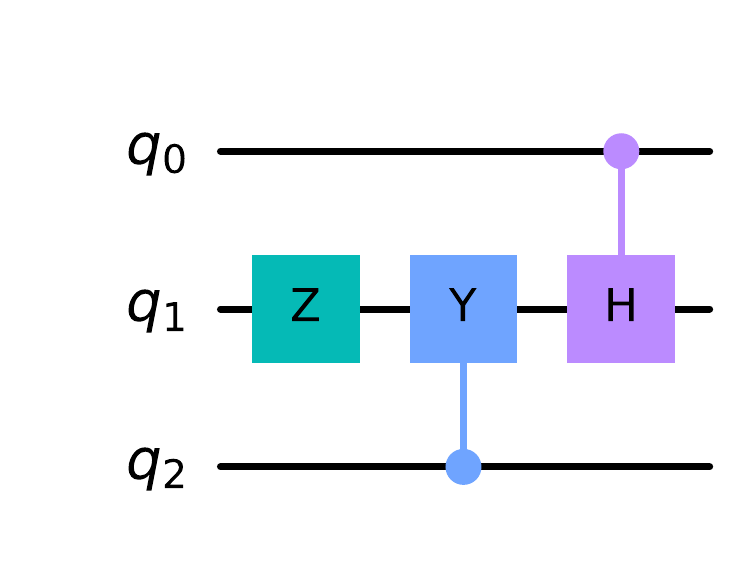}
     & \includegraphics[height=15mm,valign=m]{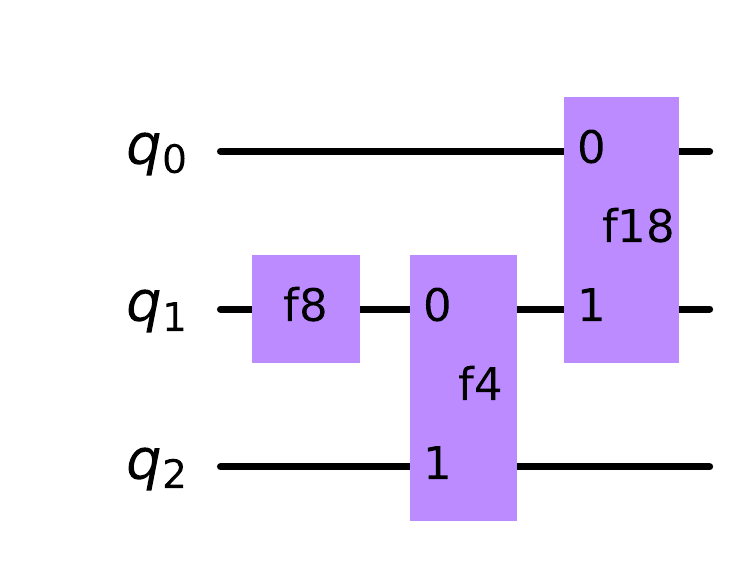}
     & \includegraphics[height=15mm,valign=m]{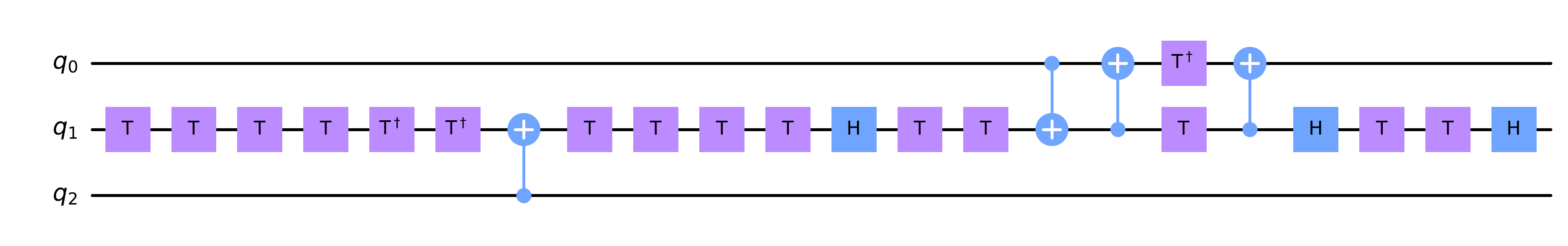}          \\
\end{longtblr}

\clearpage
\end{document}